\pdfoutput=1
\documentclass[fleqn,usenatbib,useAMS]{mnras}

\usepackage{amssymb}	% Extra maths symbols
\usepackage{multicol}        % Multi-column entries in tables
\usepackage{bm}		% Bold maths symbols, including upright Greek
\usepackage{pdflscape}	% Landscape pages
\usepackage{ae,aecompl}

\usepackage {graphicx}
\usepackage{natbib}
\usepackage{aas_journals}
\usepackage{color}
\usepackage{amsmath}
\usepackage{newtxtext}
\newcommand{\redmagic}{\textit{redMaGiC}}

\newcommand{\mpc}{h^{-1}\mathrm{Mpc}}

\DeclareMathAlphabet{\mathbbm}{U}{bbm}{m}{n}
\title[DES and the Cold Spot]{The DES view of the Eridanus supervoid and the CMB Cold Spot}

\author[A. Kov\'acs et al. (the DES Collaboration)]{
\parbox{\textwidth}{
\Large
A.~Kov\'acs$^{1,2}$\thanks{Juan de la Cierva Fellow, corresponding author: \texttt{\rm \texttt{akovacs@iac.es}}}
N.~Jeffrey$^{3,4}$,
M.~Gatti$^{5,6}$,
C.~Chang$^{7,8}$,
L.~Whiteway$^{4}$,
N.~Hamaus$^{9}$,
O.~Lahav$^{4}$,
G.~Pollina$^{9}$,
D.~Bacon,$^{10}$
T.~Kacprzak,$^{11}$
% ...
}
  \vspace{0.05cm}\\~\\
\parbox{\textwidth}{\centering \textsc{\large(The DES Collaboration)} \\ \centering \textit{Author affiliations are listed at the end of this paper}\\ }}
\begin{document}
\label{firstpage}
\maketitle

\begin{abstract}
The Cold Spot is a puzzling large-scale feature in the Cosmic Microwave Background temperature maps and its origin has been subject to active debate. As an important foreground structure at low redshift, the Eridanus supervoid was recently detected, but it was subsequently determined that, assuming the standard $\Lambda$CDM model, only about 10-20$\%$ of the observed temperature depression can be accounted for via its Integrated Sachs-Wolfe imprint. However, $R\gtrsim100~\mpc$ supervoids \emph{elsewhere} in the sky have shown ISW imprints $A_{\mathrm{ISW}}\approx5.2\pm1.6$ times stronger than expected from $\Lambda$CDM ($A_{\mathrm{ISW}}=1$), which warrants further inspection. 
Using the Year-3 redMaGiC catalogue of luminous red galaxies from the Dark Energy Survey, here we confirm the detection of the Eridanus supervoid as a significant under-density in the Cold Spot's direction at $z<0.2$. We also show, with $\mathrm{S/N}\gtrsim5$ significance, that the Eridanus supervoid appears as the most prominent large-scale under-density in the dark matter \emph{mass maps} that we reconstructed from DES Year-3 gravitational lensing data. 
While we report no significant anomalies, an interesting aspect is that the amplitude of the lensing signal from the Eridanus supervoid at the Cold Spot centre is about $30\%$ lower than expected from similar peaks found in N-body simulations based on the standard $\Lambda$CDM model with parameters $\Omega_{\rm m} = 0.279$ and $\sigma_8 = 0.82$. Overall, our results confirm the causal relation between these individually rare structures in the cosmic web and in the CMB, motivating more detailed future surveys in the Cold Spot region.
\end{abstract}


\section{Introduction}
\label{Section1}
The \emph{Cold Spot} (CS) is a large-scale anomaly of about $10^\circ$ diameter in the Cosmic Microwave Background (CMB) temperature maps. Centred on $l,b \simeq 209^\circ,-57^\circ$ galactic coordinates, it was first detected using a spherical harmonic wavelet filtering method \citep{CruzEtal2004} in Wilkinson Microwave Anisotropy Probe (WMAP) data set \citep{bennett2012}, and it was later confirmed in {\it Planck} data \citep{Planck23}. Subsequently, \cite{ZhangHuterer2010} and \cite{Nadathur2014} pointed out that the most anomalous nature of the CS is not primarily its coldness at its centre, but rather the combination of a cold interior \emph{and} a surrounding hot ring. Overall, the CS is equivalent to a $\sim$3$\sigma$ fluctuation in an ensemble of Gaussian CMB skies, thus the null hypothesis of a tail-end primordial temperature fluctuation cannot be rejected. 

\begin{table}
\centering
\caption{\label{tab:table0} A collection of the main proposed explanations for the Cold Spot and their current status in the light of existing observational probes.}
\begin{tabular}{@{}ccc}
\hline
 & {\bf Cold Spot Hypothesis}  & {\bf Observational Status} \\
\hline
1. & Measurement error & excluded, \emph{Planck} data confirmed it \\
2. & Galactic foreground & excluded, no frequency dependence \\
3. & Sunyaev-Zeldovich effect & excluded, no major low-$z$ cluster \\
4. & Cosmic texture & no evidence from other probes\\
5. & Multiverse signature & no evidence, highly speculative \\
6. & Primordial fluctuation & possible, formally a $\sim$3$\sigma$ CMB fluke \\
7. & Imprint of a supervoid & possible, anomalies in other voids  \\
8. & Combined effect from 1-7 & possible, depends on cosmology \\
\hline
\end{tabular}
\end{table}

\begin{figure}
\begin{center}
\includegraphics[width=84mm
]{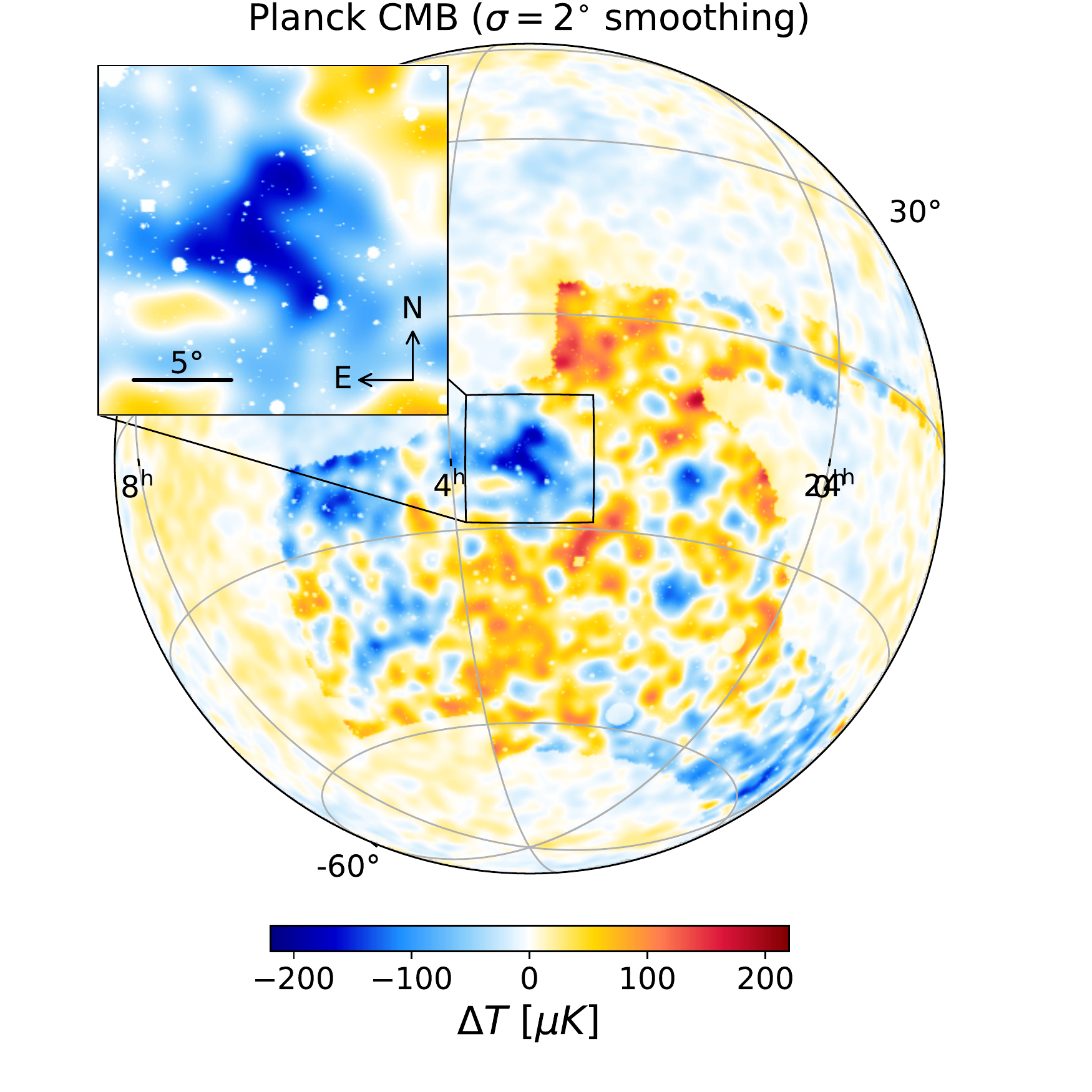}
\end{center}
\caption{The {\it Planck} CMB temperature map \citep[][]{Planck2018_cosmo} is shown centred on the CS with a Gaussian smoothing applied to suppress small-scale fluctuations (using RA and Dec Equatorial coordinates). We highlighted the DES Y3 survey footprint which is the basis of our investigations in this paper. The inset shows a $10^{\circ}\times10^{\circ}$ zoom-in version of the CS area.}
\label{fig:coldspot}
\end{figure}

Yet, there has been an active debate about possible physical processes from foreground structures, at low or high redshifts, that might imprint such a spot on the CMB sky (see Table \ref{tab:table0} for popular hypotheses). The first proposals for the physical origin of the CS included rather exotic physics, e.g., cosmic textures at $z\approx1$ \citep{CruzEtal2008}, without valuable supporting evidence from independent probes \citep[see][for a detailed review]{Vielva2010}.  
Another active line of follow-up research was focused on the possible existence of a large under-density, a \emph{supervoid}, in the matter distribution in alignment with the CS. The rationale is that, in the presence of a dominant low-$z$ dark energy component, the decaying gravitational potential ($\Phi$) of a supervoid may generate at least a significant fraction of the observed temperature depression ($\Delta T_{\rm 0} \approx -150~\mu K$ in the centre, see Figure \ref{fig:coldspot} for details) via the late-time Integrated Sachs-Wolfe (ISW) effect \citep{SachsWolfe}. In general, the total ISW shift of the CMB photon temperatures along a direction $\boldsymbol{\hat n}$ can be calculated from the time-dependent gravitational potential ($\dot{\Phi}\neq0$) based on the line-of-sight integral (other notations may use conformal time)
\begin{equation}
\label{eq:ISW_definition}
\frac{\Delta T_\rmn{ISW}}{\overline{T}}(\boldsymbol{\hat n}) = 2\int_0^{z_\rmn{LS}} \frac{a}{H(z)}\dot\Phi\left(\boldsymbol{\hat n},z\right)\,\rmn{d}z\;,
\end{equation}
with $c$=1, scale factor $a=1/(1+z)$, Hubble parameter $H(z)$, extending to the redshift of last scattering $z_\mathrm{LS}$. 
The gravitational potential $\Phi$ is related to the matter density field  $\delta(\boldsymbol{r})$ by the Poisson equation 
\begin{equation}
\label{eq:Poisson}
\nabla^2 \Phi(\boldsymbol{r},z) = \frac{3}{2}H_{\rmn{0}}^2\Omega_{\rmn{m}}\frac{\delta(\boldsymbol{r},z)}{a}\,,
\end{equation} 
with the Hubble constant $H_{\mathrm{0}}$ and matter density parameter $\Omega_{\mathrm{m}}$. In the linear growth approximation, density perturbations grow as $\delta(\boldsymbol{r},z) =  D(z)\delta(\boldsymbol{r})$, where $D(z)$ is the linear growth factor with a typical normalisation $D(0)=1$. We note that there are subdominant non-linear Rees-Sciama effects \citep[RS,][]{ReesSciama} that remain at the $\Delta T_{\rm RS} / \Delta T_{\rm ISW}\lesssim10\%$ level compared to the ISW term \citep[see e.g.][for simulated results]{Cai2010}.

From a combination of linear growth and the Poisson equation, it follows that $\dot{\Phi}=-H(z)[1-f(z)]\Phi$, where $f= \mathrm{d}\ln D /\mathrm{d}\ln a$ is the linear growth rate of structure. Then, one can obtain the following formula for the linear ISW effect that is dominant at large scales:
\begin{equation}
\label{eq:ISW_definition2}
\frac{\Delta T_\rmn{ISW}}{\overline{T}}(\boldsymbol{\hat n}) = -2\int_0^{z_\rmn{LS}} a\left[1-f(z)\right]\Phi\left(\boldsymbol{\hat n},z\right)\,\rmn{d}z\;.
\end{equation}

Considering the theoretical side of the problem, we note that the ISW signal is sensitive to the underlying cosmology \citep[see e.g.][]{CaiEtAl2014,Beck2018,Adamek2020}, and this makes the proposed causal relation of a foreground supervoid and the CS an interesting hypothesis. Since the ISW signal is sourced by a suppression of the growth rate ($f<1$) due to the extra space-stretching effects by dark energy at low redshifts, the details of the measured ISW imprints can constrain the properties of the cosmological constant ($\Lambda$) in the $\Lambda$-Cold Dark Matter ($\Lambda$CDM) model.

Note that in order to imprint a strong ISW signal, a supervoid should ideally be located at low redshift where the $[1-f(z)]$ growth suppression factor is the strongest (see Equation \ref{eq:ISW_definition2}). Earlier in the $\Lambda$CDM timeline during the Einstein-de Sitter-like matter dominated epoch at $z\gtrsim2$, gravitational growth and cosmic expansion are in balance ($f\approx1$), implying constant gravitational potentials ($\dot{\Phi}\approx0$) and $\Delta T_{\rm ISW} \approx 0$. Based on Equation \ref{eq:ISW_definition2}, the second requirement for a strong ISW imprint is a large fluctuation in $\Phi$ \citep[see e.g.][]{NadathurEtal2016}, which is also more easily met in the low-$z$ range where the cosmic web features bigger density fluctuations.

The first expectation was that an exceptionally deep ($\delta_{\rm 0} \approx -1$) and very large ($R\gtrsim 200~\mpc$) supervoid at $z\approx1$ could, in principle, imprint the CS via the ISW effect \citep{InoueSilk2006,InoueSilk2007,InoueEtal2010}. However, a sensible reasoning to exclude \emph{any} supervoid explanation was that in the standard theory of peaks for Gaussian random fields \citep{BBKS1986} the probability of the formation of a supervoid capable of imprinting a CS-like profile via the ISW effect is practically zero (corresponding to a $\gtrsim$5$\sigma$ density fluctuation) in $\Lambda$CDM \citep[][]{Nadathur2014}. The CS itself is a $\sim$3$\sigma$ fluctuation in Gaussian CMB map statistics of cold spots and thus hypothesising such an unlikely supervoid makes no sense in solving the problem itself. 

Alternatively, the CS might be composed of a moderate negative fluctuation in the primordial CMB plus a small negative $\Delta T_{\rm ISW}$ contribution, rather than it being entirely a primordial fluctuation or entirely an ISW imprint. Smaller and/or shallower voids at lower redshift may \emph{partially} contribute to the observed temperature depression, but such structures are incapable of explaining the total CS profile in a standard $\Lambda$CDM model \citep[][]{Naidoo2016}.

\begin{figure}
\begin{center}
\includegraphics[width=89mm
]{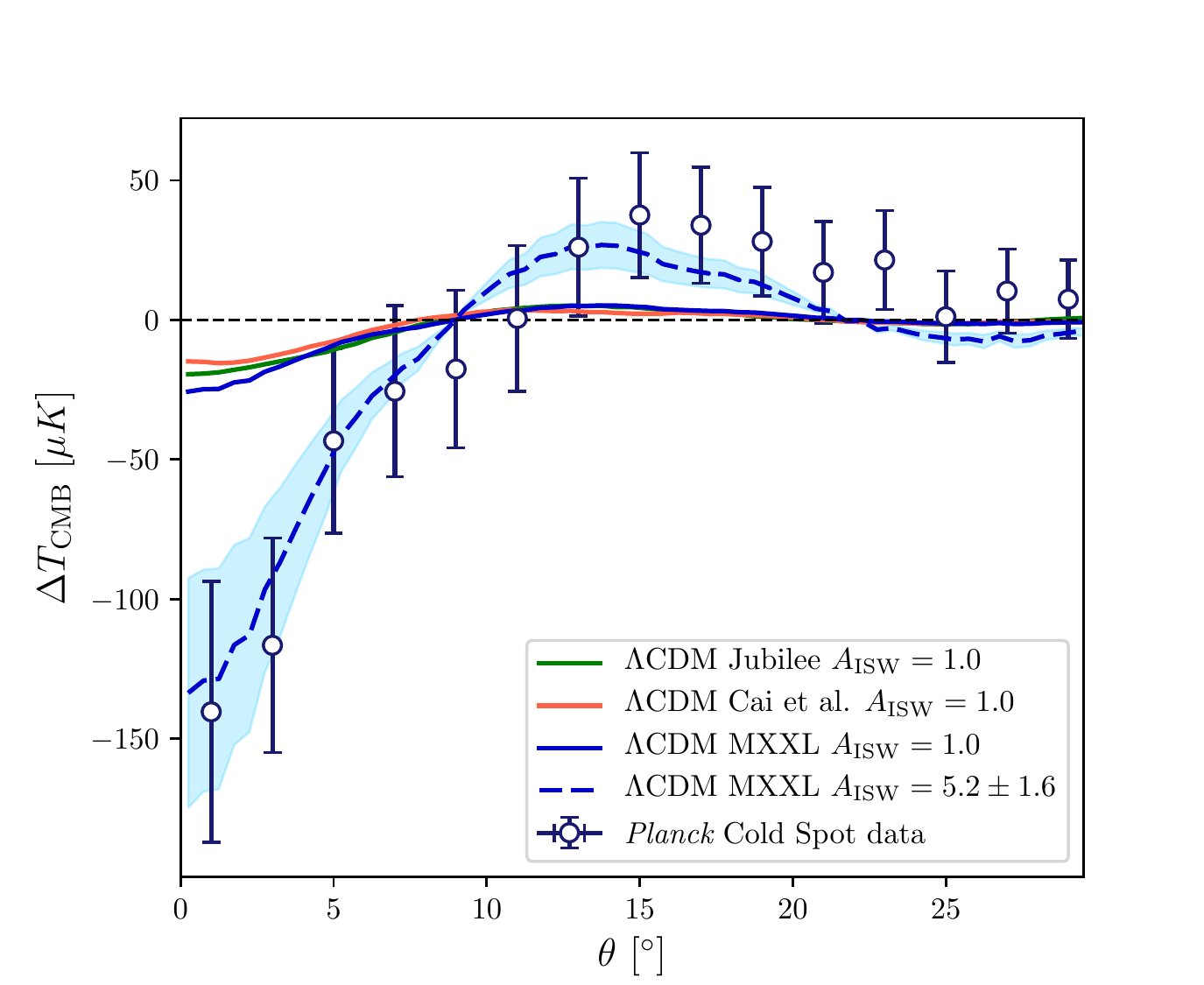}
\end{center}
\caption{A comparison of the observed CS data and the $\Delta T_{\rm ISW}$ profiles aligned with ``coldest spots'' in three different N-body simulations assuming a $\Lambda$CDM model ($A_{\rm ISW}=1$, $\Delta T_{\rm 0} \approx -20~\mu K$). For the MXXL simulation, we also show an enhanced ISW signal by \emph{re-scaling} the $\Lambda$CDM template with $A_{\rm ISW}\approx5.2\pm1.6$ that was determined from the observational analysis of several other supervoids, including DES Y3 and BOSS DR12 data.}
\label{fig:profiles_CSmodel}
\end{figure}

On the observational front, the existence of significant voids at $z\gtrsim0.3$ has been excluded with high confidence \citep{GranettEtal2010,BremerEtal2010}, in line with the theoretical expectations discussed above. As a culmination of an extensive search, the relatively shallow ($\delta_\mathrm{0}\approx-0.2$), but certainly extended ($R\approx200~\mpc$) \emph{Eridanus supervoid} was discovered at $z\approx0.15$ in the direction of the CS \citep[see e.g.][and Section \ref{sec:section_mapping} below]{SzapudiEtAl2014}. 
Assuming a baseline $\Lambda$CDM model, there is a consensus about the corresponding ISW imprint of supervoids with parameters consistent with the above observationally determined values \citep[see e.g.][]{Nadathur2014,MarcosCaballero2015,FinelliEtal2014,Naidoo2016,Naidoo2017}. As shown in Figure \ref{fig:profiles_CSmodel}, the expected central ISW imprint is of order $\Delta T_{\rm 0} \approx -20~\mu K$, in accordance with the ``coldest spot'' in the simulated Jubilee \citep[][]{Watson2014}, Millennium XXL \citep{Angulo2012}, and \cite[][]{Cai2010} ISW maps using the same definition and wavelet filtering technique on mock ISW maps \citep{Kovacs2018,Kovacs2020}. 

Therefore, among others, \cite{Nadathur2014} and \cite{Mackenzie2017} concluded that a causal relation is certainly plausible, but the Eridanus supervoid can only explain about 10-20$\%$ of the observed CS profile ($\Delta T_{\rm 0} \approx -150~\mu K$) in the $\Lambda$CDM model.

Nonetheless, an important further aspect is that the \emph{observed} amplitude of the ISW signal ($A_\mathrm{ISW}\equiv \Delta T_\mathrm{obs}/\Delta T_\mathrm{\Lambda CDM}$) is often significantly higher from supervoids than expected in the concordance model ($A_\mathrm{ISW}=1$). 
Such excess ISW signals were first found by \cite{Granett2008} using luminous red galaxies (LRG) from the Sloan Digital Sky Survey (SDSS) data set. Follow-up measurements and simulation analyses then determined that the observed signal is in about  $\sim3\sigma$ tension with the $\Lambda$CDM model expectations \citep[see e.g.][]{Nadathur2012,Flender2013,Hernandez2013,Ilic2013,Aiola}. 

To further test the claimed ISW anomalies, \cite{Kovacs2016} used photo-$z$ catalogues of LRGs from the Dark Energy Survey \citep{DES_review} Year-1 data set (DES Y1) and reported an excess signal, similar to the original SDSS detection by \cite{Granett2008}. This analysis was extended to the DES Year-3 data set and the excess ISW signals were confirmed \citep{Kovacs2019}. These findings were crucial, because they independently detected ISW anomalies using a \emph{different} part of the sky. 

In combination with the BOSS results using similarly defined supervoids \citep{Kovacs2018}, the ISW amplitude from BOSS DR12 and DES Y3 data is $A_\mathrm{ISW}\approx5.2\pm1.6$ in the $0.2<z<0.9$ redshift range. Note that this excess ISW amplitude appears to be consistent with the enhancement that would be necessary to fully explain the CS as an ISW imprint from the Eridanus supervoid, as shown in Figure \ref{fig:profiles_CSmodel}. An obvious question to ask: is this a coincidence, or the two ISW\emph{-like} anomalies concerning supervoids are related?

In this paper, we approach this problem from a different perspective. Unlike ISW measurements, recent analyses of the CMB \emph{lensing} imprints of the anomalous supervoids showed no excess signal neither using BOSS \citep[][]{Cai2017,Raghunathan2019} nor DES Y1 data \citep[][]{Vielzeuf2019}, which might provide new insights. Along similar lines, here we use the state-of-the-art dark matter \emph{mass maps} reconstructed using the DES Y3 data set \citep{Jeffrey2021} to study the gravitational lensing signal of the Eridanus supervoid. In particular, we explore how special it is in the 4100 deg$^{2}$ DES Y3 footprint and how its shape and amplitude compare to $\Lambda$CDM expectations \citep[see e.g.][]{Higuchi2018}.

The paper is organised as follows. In Section \ref{sec:section_mapping}, we provide a historical account of cosmographic analyses in the CS area, and also describe the DES data sets. We present our observational methods and results in Section \ref{sec:section_methods}, and then compare our findings to simulated results in Section \ref{sec:section_nbody} . Finally, Section \ref{sec:section_disc} presents a discussion of our main results including our conclusions.

\section{Mapping the Cold Spot area}
\label{sec:section_mapping}

\subsection{Existing results: from galaxy maps to cosmic flows}
The first evidence for an under-density aligned with the CS was presented by \cite{RudnickEtal2007} by studying a catalogue of radio galaxies in the NRAO VLA Sky Survey (NVSS). However, no redshift information was available for the void candidate, and the significance of the detection was disputed \citep{SmithHuterer2010}. 

Targeted pencil beam surveys \cite{GranettEtal2010} and \cite{BremerEtal2010} found no evidence for a significant under-density between redshifts of $0.5<z<0.9$, but their galaxy counts were consistent with a void at $z<0.3$. In addition, the analysis of the 2-Micron All-Sky Survey Extended Source Catalog \citep[2MASS XSC]{jarrett2000} galaxy distribution by \cite{francis2010} showed a shallow under-density of large angular size around the CS. \cite{rassat2013} confirmed the presence of this low-$z$ void in the reconstructed 2MASS ISW maps. \cite{Manzotti2014} found that any late time ISW-RS imprints that might be responsible for the CS are very likely to originate at $z < 0.3$, motivating a detailed examination of this range. 

Then, \cite{FinelliEtal2014} analysed the low-$z$ WISE-2MASS galaxy catalogue \citep{KovacsSzapudi2014} that combines measurements of two all-sky surveys in the infrared, the Wide-Field Infrared Survey Explorer \citep[WISE]{wise} and the Point Source Catalog of the 2-Micron All-Sky Survey \citep[2MASS]{2mass}. They identified a $\sim 20\%$ under-density ($\delta_{\rm 0} \approx -0.2$) in the direction of the CS with $\theta\approx20^{\circ}$ angular size that corresponds to about $R\approx200~\mpc$ physical radius, i.e. a rare density fluctuation given the combination of size and under-density. 

Along similar lines, \cite{SzapudiEtAl2014} matched the WISE-2MASS galaxy data set to a 1,300 deg$^2$ area with Pan-STARRS1 \citep[PS1]{ps1ref} data, adding optical colours for each object. For the resulting catalogue, photometric redshifts were estimated and the line-of-sight galaxy density profile was analysed in the redshift range $ z < 0.3$. Further evidence was found for a shallow but extended supervoid ($\delta_{\rm 0} \approx -0.2$, $R\approx200\mpc$) centred on the CS, indicating a roughly spherical structure in combination with the constraints on it transverse size by \cite{FinelliEtal2014}. 

After the discovery of this under-density called the \emph{Eridanus supervoid}, \cite{KovacsJGB2015} analysed the 2MASS photo-$z$ catalogue \citep[][2MPZ]{Bilicki2014} and found that the supervoid is elongated and its extent in the line-of-sight might be larger than $R\approx200~\mpc$, extending to the lowest redshifts.

A further development was the dedicated 2CSz spec-$z$ survey of about 7000 galaxies in the CS region at $z<0.4$. \cite{Mackenzie2017} identified 4 smaller voids at different redshifts which suggests sub-structure for the Eridanus supervoid. However, they also identified a similar under-density in the line-of-sight in their control sample elsewhere in the sky (not observed by DES). They then argued that the absence of a CS-like pattern in alignment with this other system of voids suggests that the Eridanus supervoid is not a special structure in the low-$z$ Universe and therefore there is no causal relation with the CS. Alternatively, the overall \emph{volume} of the Eridanus supervoid in 3D may be larger than that of the under-density in the control field by \cite{Mackenzie2017} depending on the large-scale environments around the measured lines of sight. This second hypothesis is also supported by \cite{Courtois2017} who analysed their Cosmicflows-3 data set and found that the ``Cold Spot Repeller'' is the largest basin of repulsion in the $z\approx0.1$ cosmic web, closely aligned with the CS. Therefore, the Eridanus supervoid appears to be a rare under-density in a full 3D view and further investigations are needed to determine its connections to the CS; including this analysis using DES data.

\subsection{Motivation: the lensing imprint of voids}

The weak lensing information from DES Y3 data is a key novelty in the problem of the CS and the Eridanus supervoid. In a wider context, it helps to contribute to an established line of research on the mass distribution in cosmic voids, including possible precision tests of alternative cosmological models in void environments \citep[see e.g.][]{Clampitt2013,Cai2015,Cautun2018,Baker2018,Davies2020}. 

Unlike clusters, groups, and filaments, cosmic voids cause a de-magnification effect and therefore correspond to local minima in the lensing convergence ($\kappa$) maps, estimated from the matter density field $\delta(r,\theta)$ via projection as
\begin{equation}
    \kappa(\theta)=\frac{3H_0^2\Omega_m}{2c^2}
    \int_{0}^{r_{\rm max}} \delta (r,\theta)
    \frac{(r_{\rm max}-r)r}{r_{\rm max}}\, dr,
    \label{eq:kappa_born}
\end{equation}
where $r$ denotes a co-moving distance to the sources, and $r_{\rm max}$ determines the maximum distance considered in the projection.

Several real-world detections of the lensing signal from multiple voids using stacking methods have already been reported \citep{Melchior2014,Sanchez2016,Gruen2016,ClampittJain2015,Brouwer2018,Fang2019}, including CMB lensing analyses \citep{Cai2017,Vielzeuf2019,Raghunathan2019,Hang2021}. These observations are supported and complemented by signal-to-noise optimisation efforts and tests of the signal shape given different void  definitions using N-body simulations \citep[see e.g.][]{Cautun2016,Davies2018,Davies2021}.

However, the detection of the weak lensing effect of an \emph{individual} void has been considered a great challenge due to the significant measurement uncertainties \citep[see e.g.][]{Krause2013}, except in the case of the largest voids with $R\gtrsim100~\mpc$ radius \citep{Amendola1999}. Therefore, the Eridanus supervoid is a good candidate for such a measurement given its low redshift and approximate $R\approx200~\mpc$ radius. 

\begin{figure*}
\begin{center}
\includegraphics[width=178mm
]{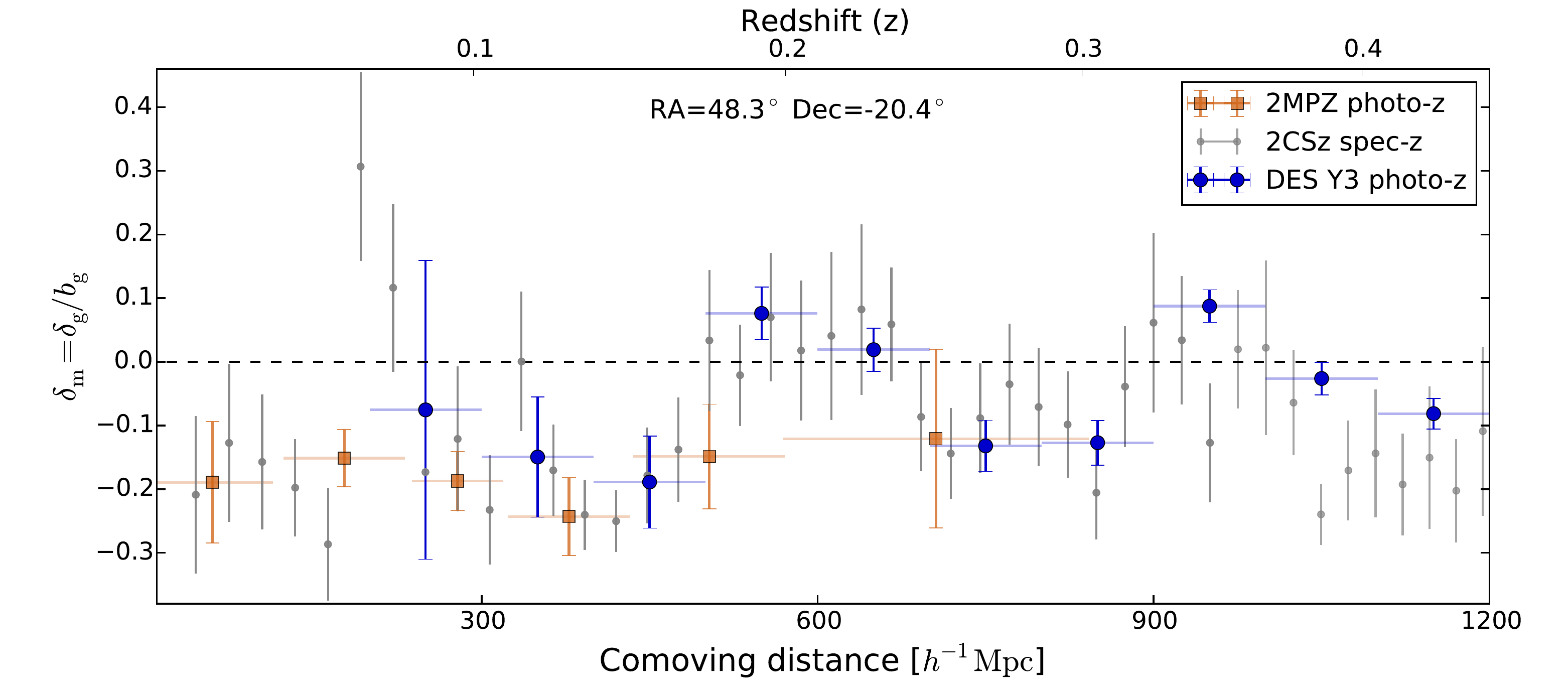}
\end{center}
\caption{The line-of-sight matter density profiles are compared for different surveys of the CS direction. We converted galaxy density to matter density using the independently determined linear galaxy bias ($b_{\rm g}$) values for each tracer data set. The DES results from \redmagic{} galaxies feature a consistent density profile when compared to 2MPZ (photo-$z$) and 2CSz (spec-$z$) data. We also observe two previously seen smaller voids at redshifts higher than the main supervoid at $z<0.2$, which may also contribute to the overall ISW imprints of the supervoid.}
\label{fig:profile_CS}
\end{figure*}

\cite{Higuchi2018} used an N-body simulation with a lensing convergence map and found an expected significance of $\mathrm{S/N}\gtrsim 4$ for a weak lensing signal from such an extended low-$z$ under-density assuming a standard $\Lambda$CDM model. \cite{Higuchi2019} also showed that measurable decreasing trends are expected in the non-Gaussian peak statistics in $\kappa$ maps towards the largest under-dense regions such as the Eridanus supervoid. These recent findings from simulations provide motivation for our observational analysis. Our main methods are the following:
\begin{itemize}
    \item we first measure a \emph{line-of-sight} galaxy density profile at the CS using LRGs selected from DES data.
    \item we then rely on the gravitational lensing convergence ($\kappa$) maps, reconstructed from DES cosmic shear measurements, and follow up on the detection of the Eridanus supervoid using dark matter mass maps, and compare the results to N-body simulations. 
\end{itemize}

\subsection{DES Y3 data: Luminous red galaxies}

We mapped the CS region using data products from the first three years (Y3) of the Dark Energy Survey \citep[DES,][]{DES_review,DES_DR1}. DES is a six-year survey that covers approximately $4100~\mathrm{\deg}^2$ sky area of the South Galactic Cap. Mounted on the Cerro Tololo Inter-American Observatory (CTIO) four metre Blanco telescope in Chile, the $570$~megapixel Dark Energy Camera \citep[DECam,][]{Flaugher2015} images the field in $grizY$ filters. 

The raw images were processed by the DES Data Management (DESDM) team  \citep{Sevilla2011,Morganson2018}. We adopted the empirically constructed DES Y3 survey mask in our analysis, which excludes potentially contaminated pixels e.g. in the close proximity of bright stars. For the full details of the DES Y3 data set, we refer the readers to \cite{y3-gold}.

To estimate the line-of-sight galaxy density profile aligned with the CS, we used an LRG sample from the first three years of observations. This red-sequence MAtched-filter Galaxy Catalog \cite[\redmagic,][]{Rozo2015} is a catalogue of photometrically selected LRGs, based on the red-sequence MAtched-filter Probabilistic Percolation (redMaPPer) cluster finder algorithm \citep{Rykoff2014}. We utilised the \redmagic\ sample that spans the $0.2<z<0.7$ range, because of its exquisite photometric redshifts, namely $\sigma_z/(1+z)\approx 0.02$, and a $4\sigma$ redshift outlier rate of $r_\mathrm{out}\simeq1.41\%$. The resulting galaxy sample has an approximately constant co-moving space density $\bar{n}\approx 10^{-3}h^{3}$ $\mathrm{Mpc^{-3}}$ (high density sample, brighter than 0.5$L_{*}$). 

The great photo-$z$ precision allows an accurate and robust reconstruction of cosmic void environments, and such a \redmagic\ galaxy sample has been used in a series of DES void analyses including weak lensing and ISW measurements \citep[see e.g.][]{Sanchez2016,Kovacs2016,Vielzeuf2019,Fang2019}. 

The galaxy clustering properties of the latest Y3 \redmagic\ data set are presented by \cite{Pandey2021}. In the context of possible remnant systematic effects in this sample \citep[see e.g.][for further details]{descollaboration2021dark}, we note that 3,222 DES voids, identified from the Y3 \redmagic{} data, did show correlations with DES mass map features \citep{Jeffrey2021}. This indicates that genuine cosmic voids are detected from this catalogue.

\subsection{DES Y3 data: dark matter mass maps}

Complementing galaxy catalogues, the mass maps are weighted projections of the density field (primarily dark matter) in the foreground of the observed galaxies. Following the DES Y3 mass map reconstruction analysis presented in detail by \cite{Jeffrey2021}, we consider mass maps in \texttt{HEALPix} format with resolution $N_\mathrm{side} = 1024$ \citep[][]{healpix} based on four slightly different reconstruction methods; each is a \emph{maximum a posteriori} estimate with a different model for the prior probability of the map:
\begin{itemize}
    \item The first method considered is the direct inversion of the shear field, also known as the \emph{Kaiser-Squires (KS)} method \citep[][]{Kaiser1993}, followed by a smoothing of small angular scales. 
    \item The second method uses a prior on the B-modes of the map, imposing that the reconstructed convergence field must be purely an E-mode map (\emph{null B-mode} prior); this method also includes smoothing at small scales. 
    \item The third method, the \emph{Wiener filter}, uses a Gaussian prior distribution for the underlying convergence field.
    \item  Lastly, the \texttt{GLIMPSE} method implements a sparsity prior in wavelet (starlet) space \citep[][]{Lanusse2016}, which can be interpreted as a physical model where the matter field is composed of a superposition of spherically symmetric halos.
\end{itemize}

All methods are implemented on the celestial sphere to accommodate the large sky coverage of the DES Y3 data. The mass maps were reconstructed using the DES Y3 shear catalogue \cite{y3-shapecatalog} of 100,204,026 galaxies in 4143~deg$^2$, building upon the Y3 Gold catalogue \citep{y3-gold} and using the \texttt{METACALIBRATION} algorithm \citep{HuffMcal2017, SheldonMcal2017}, which infers the galaxy ellipticities starting from noisy images of the detected objects in the \textit{r, i, z} bands.

Map-level tests against various systematics regarding the parent DES galaxy catalogues showed no significant remnant contamination. Focusing on large scales, here we take the DES Y3 mass maps as inputs and present an additional application in cosmographic analyses \citep[see][for further details about reconstruction methods and mass map properties]{Jeffrey2018,Jeffrey2021}.

\begin{figure*}
\begin{center}
\includegraphics[width=58mm
]{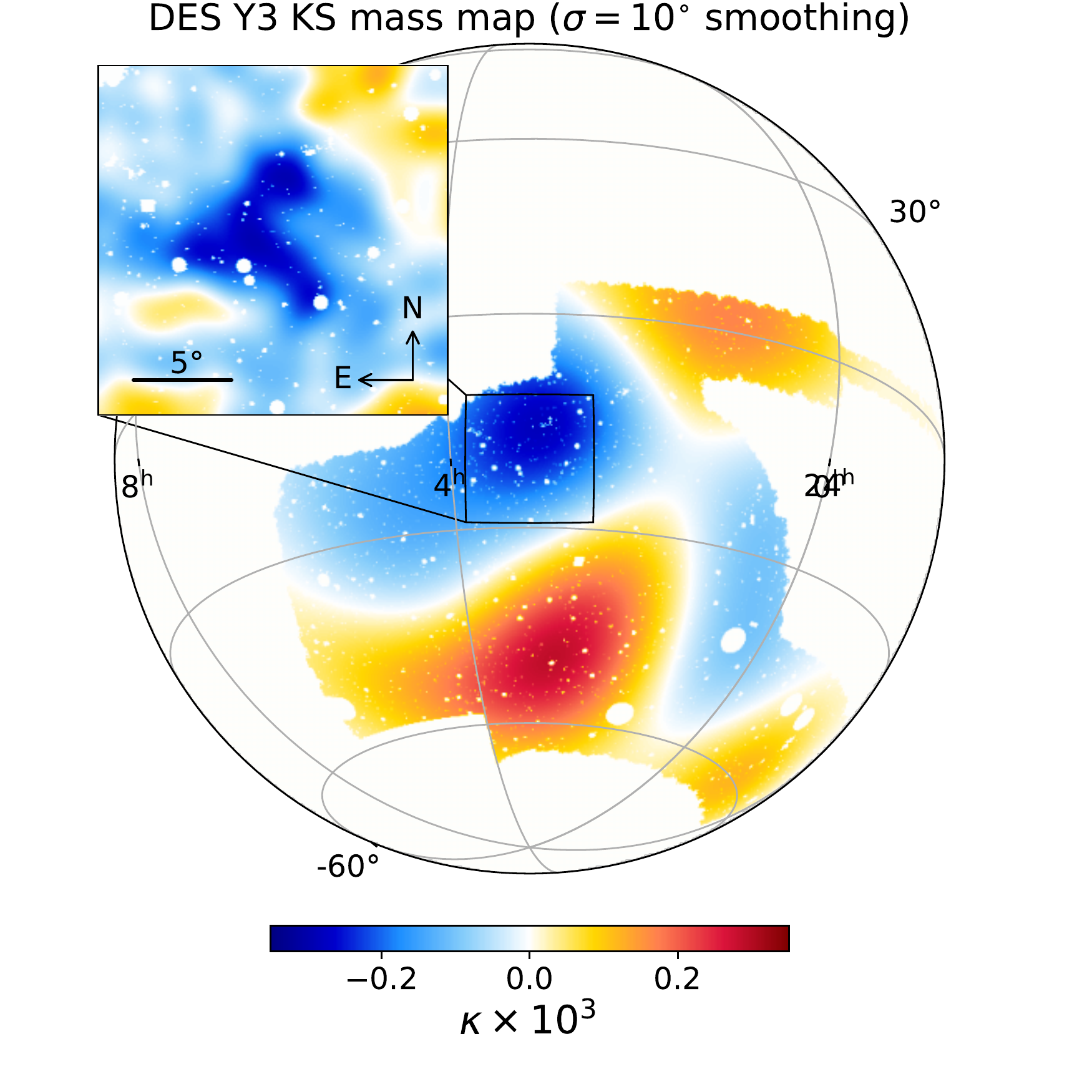}
\includegraphics[width=58mm
]{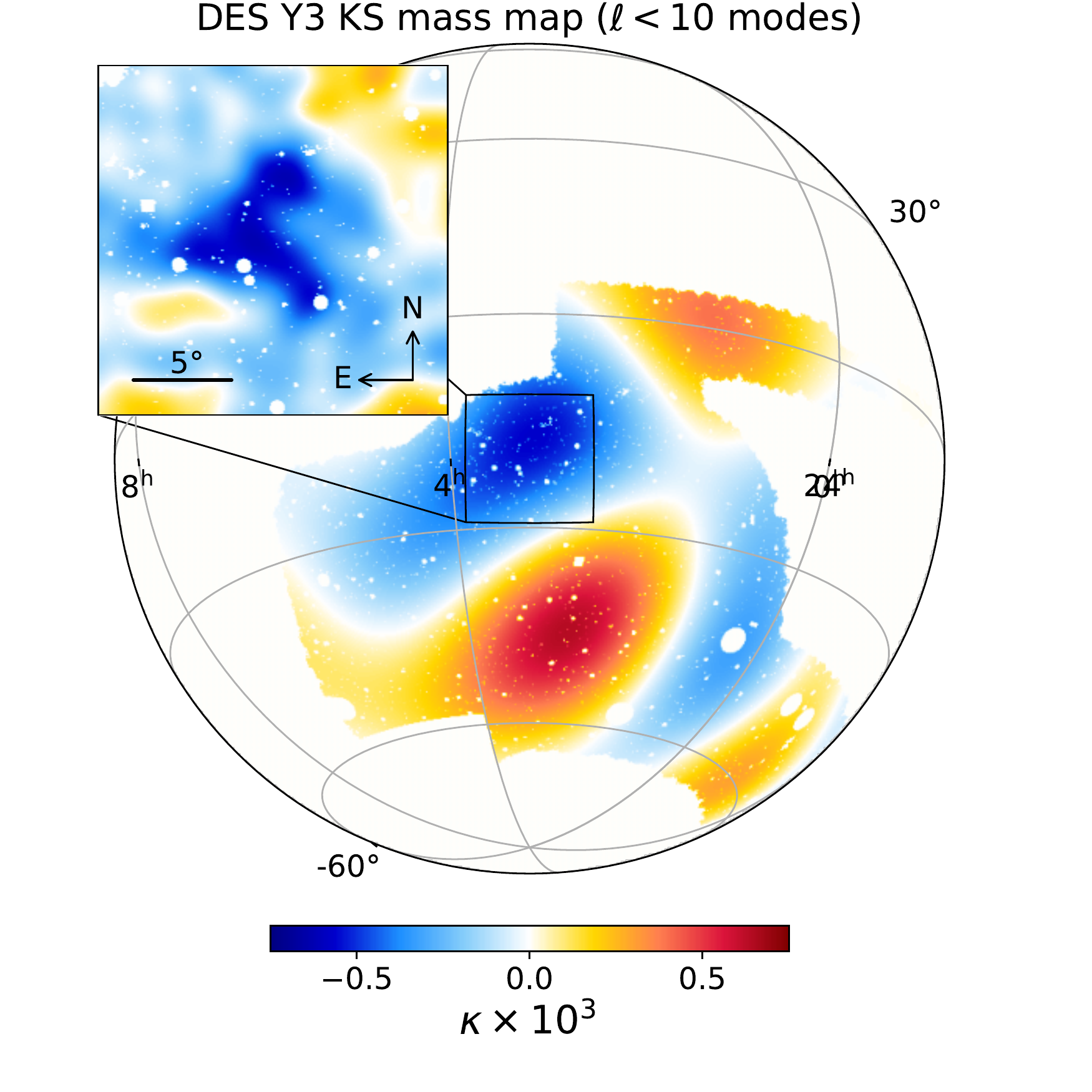}
\includegraphics[width=58mm
]{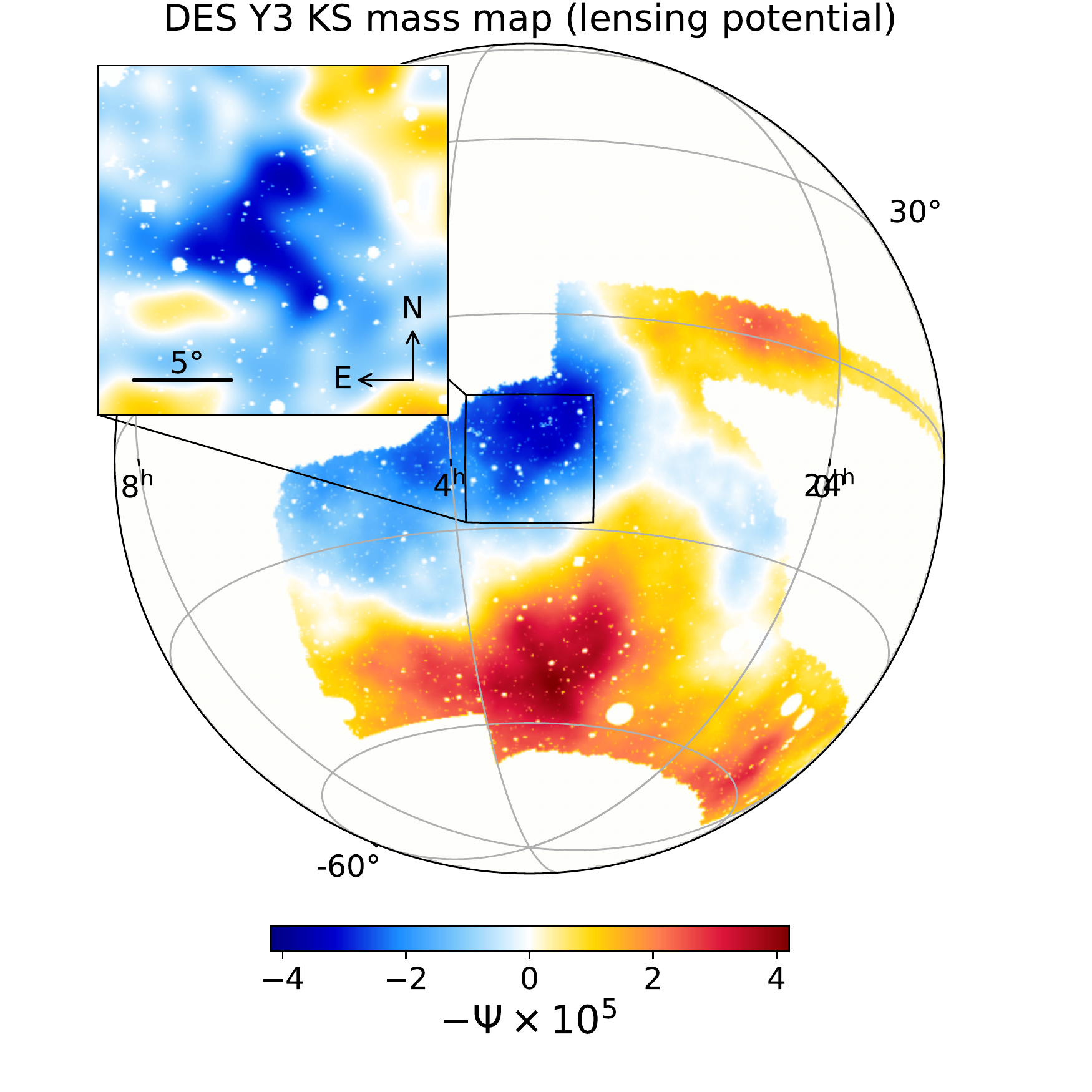}
\end{center}
\caption{Using \texttt{HEALPix} tools, the left panel shows the DES KS mass map with a $\sigma=10^{\circ}$ Gaussian smoothing applied to highlight fluctuations at the largest scales. The middle panel displays our results for a mass map with $\ell<10$ modes only. On the right, we show the projected gravitational potential ($\psi$) that we calculated from the $\kappa$ map. Our maps consistently show that the largest under-density in the DES Y3 data set is closely aligned with the CS, corresponding to a prominent low-$z$ supervoid. As in Figure \ref{fig:coldspot}, the inset in all panels shows the {\it Planck} CMB temperature map in the CS region.}
\label{fig:image_CS}
\end{figure*}

\section{Methods \& Results}
\label{sec:section_methods}
\subsection{A line-of-sight density profile}

Without considering lensing information, this first part of our analysis aims at measuring and testing the galaxy density field from the DES data in the direction of the CS, and test the consistency with previous state-of-the-art results \citep[see e.g.][]{KovacsJGB2015,Mackenzie2017}.
Following similar DES analyses presented by \cite{Chang2018} and \cite{Jeffrey2021}, the \redmagic{} galaxy catalogue was projected into two-dimensional slices of $\Delta r=100\mpc$ along the line of sight. This thickness corresponds to the approximate photo-$z$ errors of the \redmagic{} galaxies that allows the robust identification of voids \citep[see][for details]{Sanchez2016}. 

At $0.07<z<0.45$, galaxy density contrasts ($\rho_\mathrm{g}/\bar{\rho}_\mathrm{g}-1=\delta_\mathrm{g}$) are measured in 10 tomographic slices aligned with the CS at $\textrm{RA}, \textrm{Dec} \approx 48.3^{\circ}, -20.4^{\circ}$. Galaxies were counted within an aperture of $2.5^{\circ}$ of the void centre. This corresponds to approximately half the angular size of the CS, allowing a more direct analysis of the deepest parts of the supervoid. 

To facilitate comparisons to the previous 2MPZ and 2CSz surveys in the CS area that used different galaxy tracers with different linear bias, we converted the galaxy density value to matter density ($\delta_{\rm m}=\delta_{\rm g}/b_{\rm g}$) in all cases. We note that the large and shallow (super)voids detected by photo-$z$ surveys such as DES are well described by the linear bias model; they naturally trace the largest scales where the linear bias approximation is expected to hold \citep[see][for previous simulated and DES results]{Pollina2017,Pollina2019}.

Estimated from a combination of clustering and lensing 2-point correlation functions \citep{Prat2021}, the linear galaxy bias for the DES Y3 \redmagic{} sample is about $b_{\rm g}\approx1.74\pm0.12$ at our redshifts of interests (a slight redshift dependence is neglected in this analysis). The linear galaxy bias values for the 2MPZ galaxy catalogue \citep{Bilicki2014} are $b_{\rm g}\approx1.18\pm0.03$ at $z<0.08$ and $b_{\rm g}\approx1.52\pm0.03$ at $0.08<z<0.3$, as determined by \cite{Alonso2015}. The linear bias of the 2CSz galaxies changes gradually from $b_{\rm g}\approx1.35$ at $z\approx0.1$ to about $b_{\rm g}\approx2.0$ for galaxies approaching $z\approx0.4$ \citep{Mackenzie2017}. We linearly interpolated between these measured $b_{\rm g}$ values, and used them for our conversion from galaxy to matter density fluctuations.

As shown in Figure \ref{fig:profile_CS}, the matter density contrasts estimated from DES data at the different redshifts are remarkably consistent not only with the 2MPZ photo-$z$ data in their overlapping range, but also with the presumably more accurate 2CSz spec-$z$ results. This agreement further suggests that our analysis is certainly not affected by possible remnant systematic problems concerning the DES Y3 \redmagic{} catalogue \citep[][]{descollaboration2021dark}.

With the DES data, we confirmed the previously reported deepest part ($\delta\approx-0.25$) of the Eridanus supervoid at $z\approx0.15$, observed a slight over-density ($\delta\lesssim0.1$) at $z\approx0.2$, and also detected two smaller voids at higher redshifts. Our \redmagic\ LRG data contain few galaxies in the $z<0.1$ range and therefore we cannot provide a detailed comparison with existing measurements. In this generally under-dense low-$z$ environment (see 2MPZ data points), the intervening over-dense structure at $z\approx0.06$ (detected only in more accurate spec-$z$ data) is one of the outer filamentary features rooted in the Horologium supercluster, as noted previously by \cite{KovacsJGB2015} in their cosmographical analysis.

Considering the $z>0.45$ redshift range, we note that previous line-of-sight analyses of galaxy counts by \cite{GranettEtal2010} and \cite{BremerEtal2010} excluded the possibility of significant voids aligned with the CS (while their results were consistent with an under-density at $z<0.3$ where later the Eridanus supervoid was detected). For completeness, we did check the $0.45<z<0.7$ range using our DES \redmagic{} galaxies and, as expected, found no significant voids or superclusters (see Appendix \ref{appendix1} for details).

Our reconstruction of the matter density field at the CS from the DES Y3 data set confirms that the Eridnaus supervoid is among the largest known under-densities in the observable Universe. A deeper understanding of these large-scale structures is of great interest in cosmology \citep[see e.g.][]{Shimakawa2021}.

\subsection{Mass map filtering strategies}

The DES Y3 weak lensing convergence ($\kappa$) maps are exceptionally rich sources of cosmological information. For instance, additional information on the growth of structure and the clumpiness of the matter distribution might be extracted by analysing their higher order moments \citep{Gatti2020}, or their non-Gaussian peak statistics may also be measured with machine learning techniques to complement traditional 2-point function analyses \citep[see e.g.][]{Zuercher2021, Ribli2019}.

In this analysis, we are guided to ignore the information encoded in their small-scale patterns, and rather test their largest scales for three reasons. First, the existing evidence on the dimensions of the Eridanus supervoid suggests that its angular size is about $\theta\approx20^{\circ}$ as a consequence of its very low redshift and large physical size with about $R\approx200~\mpc$. Second, to test previous claims that this supervoid is not a particularly special under-density in the low-$z$ Universe \citep[see e.g.][]{Mackenzie2017} we wish to know if this region is of special significance considering the total 4100 deg$^{2}$ DES Y3 survey area. 

Third, we expect from the Poisson equation, conveniently expressed in Fourier space as $\Phi(\boldsymbol{k},z) \sim \delta(\boldsymbol{k},z)/\boldsymbol{k}^2$ using a wave vector $\boldsymbol{k}$, that the characteristic fluctuations in the gravitational potential appear on much larger scales than in the density field due to the $\boldsymbol{k}^{-2}$ factor. Since these perturbations in $\Phi$ are the actual sources of the ISW signal, it is important to focus on their reconstruction.

Along these lines, we followed three similar strategies to highlight and probe the largest scales in the DES Y3 mass maps:
\begin{itemize}
    \item as the simplest proxy for tracing large-scale patterns in the gravitational potential, we applied a Gaussian smoothing to the mass maps with $\sigma=10^{\circ}$.
    \item as an alternative method, we filtered the mass maps in spherical harmonic space and kept only the $\ell<10$ modes which correspond to the largest angular scales.
    \item we calculated the projected gravitational potential $\psi$ from the $\kappa$ maps following a $\kappa_{\ell m}=-\frac{1}{2}\ell(\ell+1)\psi_{\ell m}$ transformation which also effectively highlights the largest scales in the map.
\end{itemize}

Our reconstructed large-scale patterns in the DES Y3 KS mass map are shown in Figure \ref{fig:image_CS}. We report a striking visual correlation between the location of the CS and the largest negative density fluctuation in the maps, regardless of the methodology to probe the largest scales. We also observed a generally over-dense environment further away from the CS region. The significance of these correlations is estimated from more detailed analysis below. 

Overall, this finding certainly confirms previous detections of the Eridanus supervoid from an independent new tracer of large-scale structure. Moreover, it also indicates that this supervoid is indeed a special under-density in the low-$z$ cosmic web, and reinforces the hypothesis of causal correlations with the CS.

\subsection{Void lensing profiles at the Cold Spot}

An important consideration is that the CS is located close to the edge of the DES survey area as shown in Figure \ref{fig:image_CS}. We note that the originally designed DES footprint was in fact modified (during the survey planning stage) to fully include the CS area. Therefore, the observed DES Y3 data set does include a complete $\theta\lesssim15^{\circ}$ disk around the nominal CS centre ($\textrm{RA}, \textrm{Dec} \approx 48.3^{\circ}, -20.4^{\circ}$) which allows a detailed view of the large-scale structure of DES galaxies in this direction. However, this limitation also means that any radially averaged measurement beyond $\theta\approx15^{\circ}$ will necessarily result in a less complete reconstruction due to an increasing fraction of masked pixels beyond the edge of the survey (most importantly in the direction of the top-left corner of the insets in Figure \ref{fig:image_CS}). 

We quantified the visually compelling correlations seen in Figure \ref{fig:image_CS} by measuring the tangential $\kappa$ and $\psi$ profiles in $\Delta \theta = 2^{\circ}$ bins centred on the CS. We estimated the uncertainties of this detection by measuring the radial $\kappa$ profile in 200 random locations in the DES Y3 mass maps. The resulting covariance matrix C showed strong off-diagonal contributions, especially at the innermost 4-5 bins, which we take into account in our analysis. We then evaluated a $\chi^{2}$ statistic and calculated the signal-to-noise ratio compared to a null detection as $\mathrm{S/N}=\sqrt{\chi_\mathrm{0}^{2}-N_\mathrm{bins}}$ with $\chi_{0}^{2}=\kappa_\mathrm{data}C^{-1}\kappa_\mathrm{data}$, where $C^{-1}$ is the inverse of the covariance matrix from our randoms and $\kappa_\mathrm{data}$ are the measured data points in the radial profile. 

We compared our results from different versions of the DES Y3 mass map based on the Kaiser-Squires (KS), Wiener filter, and Null-B reconstruction methods. We note that, by construction, the \texttt{GLIMPSE} method is not adequate to accurately recover the largest scales and therefore we do not use the related mass map in our main analyses. For completeness, we nevertheless detected an under-density aligned with the CS in the \texttt{GLIMPSE} map as well, although with a lower amplitude than in the other three versions.

\begin{figure}
\begin{center}
\includegraphics[width=84mm
]{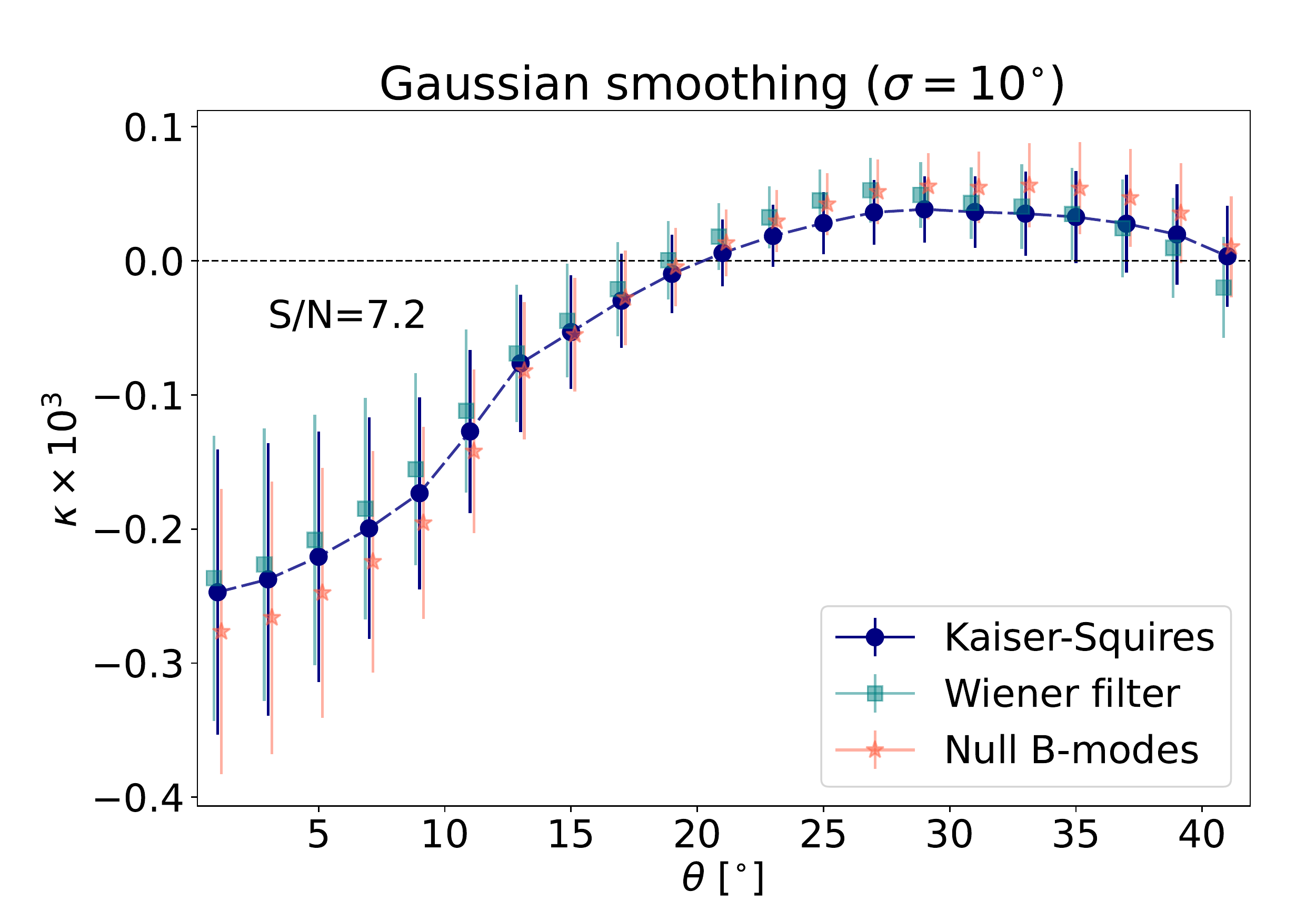}
\end{center}
\caption{Tangential $\kappa$ profiles centred on the CS. In the case of the KS map, the detection significance of the measured $\kappa$ profile is $\mathrm{S/N}\approx7.2$ compared to a null signal. We compare our 3 different mass map versions and report good consistency among them at all radii.}
\label{fig:profiles_smth}
\end{figure}

\begin{figure}
\begin{center}
\includegraphics[width=84mm
]{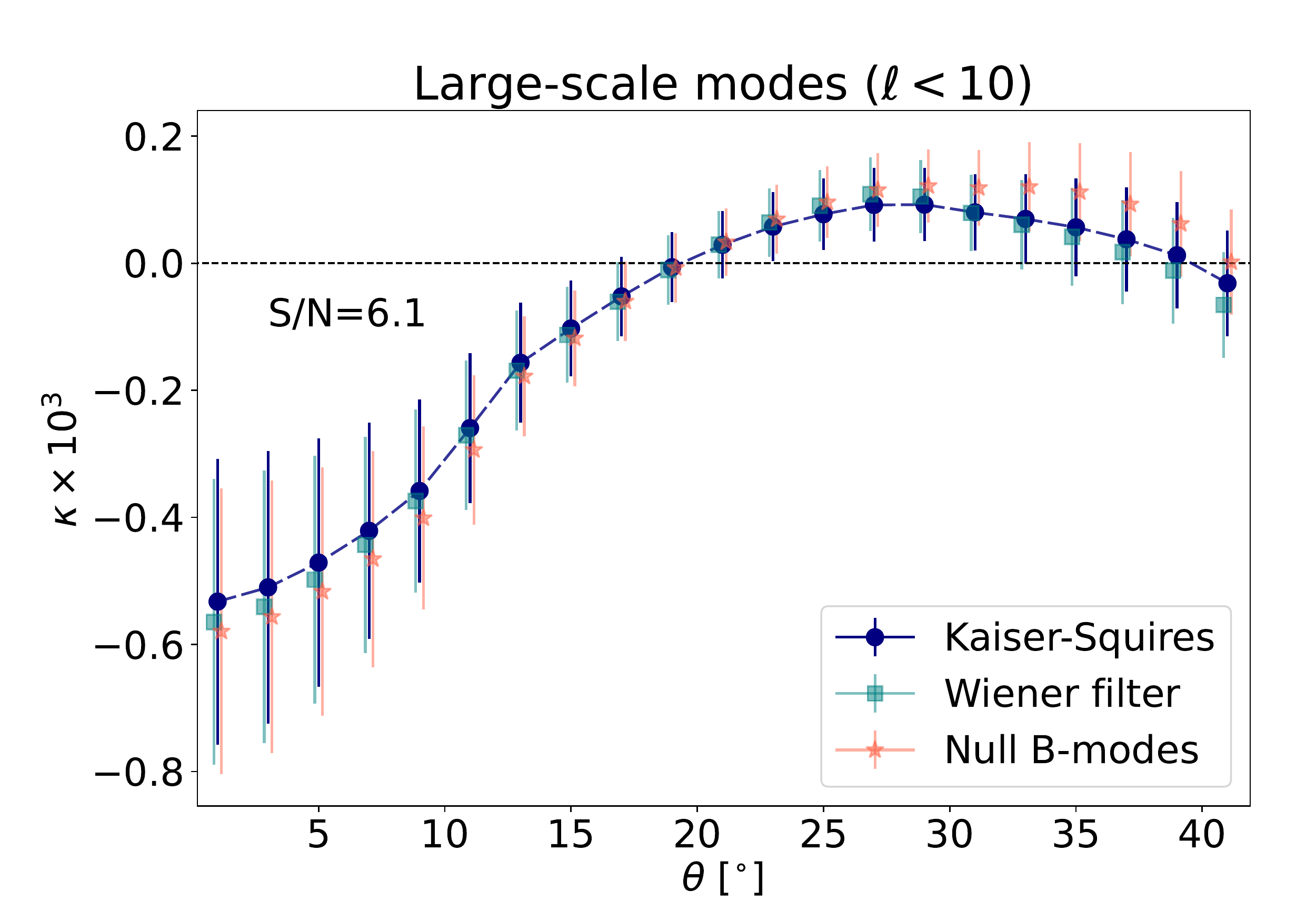}
\end{center}
\caption{Convergence ($\kappa$) profiles measured from the $\ell<10$ filtered version of the maps to study the actual large-scale modes. When comparing different mass maps, we again see good consistency throughout the full extent of the profile, including the compensation zone at $\theta \gtrsim20^{\circ}$.}
\label{fig:profiles_lowell}
\end{figure}

\subsubsection{Gaussian smoothing ($\sigma=10^{\circ}$)}

In Figure \ref{fig:profiles_smth}, we present our results in the case of the Gaussian smoothing applied to the mass maps. All maps show a consistently negative and gradually fading $\kappa$ imprint which extends to about $\theta \approx20^{\circ}$, in line with previous findings from the WISE-2MASS galaxy map \citep{FinelliEtal2014}. Testing the relation of fluctuations in mass and light, we also measured the projected galaxy density field at the CS using the DES \redmagic{} galaxies. We found an approximately linear relationship between $\delta_{\rm g}$ and $\kappa_{\rm g}$, as expected based on previous results \citep[][]{Fang2019,Pollina2019}. 

Considering only the innermost radial bin at $\theta < 2^{\circ}$, we found $\mathrm{S/N}\approx2.1$ as a signal-to-noise ratio. Using the data from all radial bins up to $\theta \approx 40^{\circ}$, we report a $\mathrm{S/N}\approx7.2$ detection of an under-density in the fiducial Kaiser-Squires mass map. While this $\mathrm{S/N}$ analysis was based on the full covariance matrix C (including bin-to-bin correlations), the error bars shown in Figure \ref{fig:profiles_smth} were estimated from the diagonal elements for demonstration. Given these uncertainties, the Wiener filtered and Null B-mode results are fully consistent with the KS signal, which strengthens our detection.

We note that a slightly over-dense ``compensation'' region ($\kappa>0$) is also detected around the central under-density ($\kappa<0$). These data points also contribute to the $\mathrm{S/N}$ that we estimate considering their covariance with the inner part of the profile. At large $\theta$, the lensing signal does approach zero as expected. See also Figure \ref{fig:image_CS} for a visual impression of the surroundings of the CS area.

\subsubsection{Large-scale modes ($\ell<10$)}

In Figure \ref{fig:profiles_lowell}, we then compare the shape of the $\kappa$ profile in the different DES mass map versions if only the largest scales are considered with $\ell<10$. The detection significance is $\mathrm{S/N}\approx2.1$ for the first bin at the centre, whilst the full profile yields a $\mathrm{S/N}\approx6.1$ detection. 

Given the errors, we again report great consistency between different mass map versions, and also with the results from Gaussian smoothing. The under-dense central part, the over-dense compensation region, and the convergence to zero signal are again clearly detected using these differently filtered maps.

We note that the amplitude of the $\kappa$ profiles from the $\ell<10$ maps are higher compared to the results from Gaussian smoothing. This is the consequence of the slightly different mass map filtering techniques that we applied to highlight the largest scales, but both sets of Wiener, Null B-mode, and KS maps are internally consistent.

\begin{figure}
\begin{center}
\includegraphics[width=84mm
]{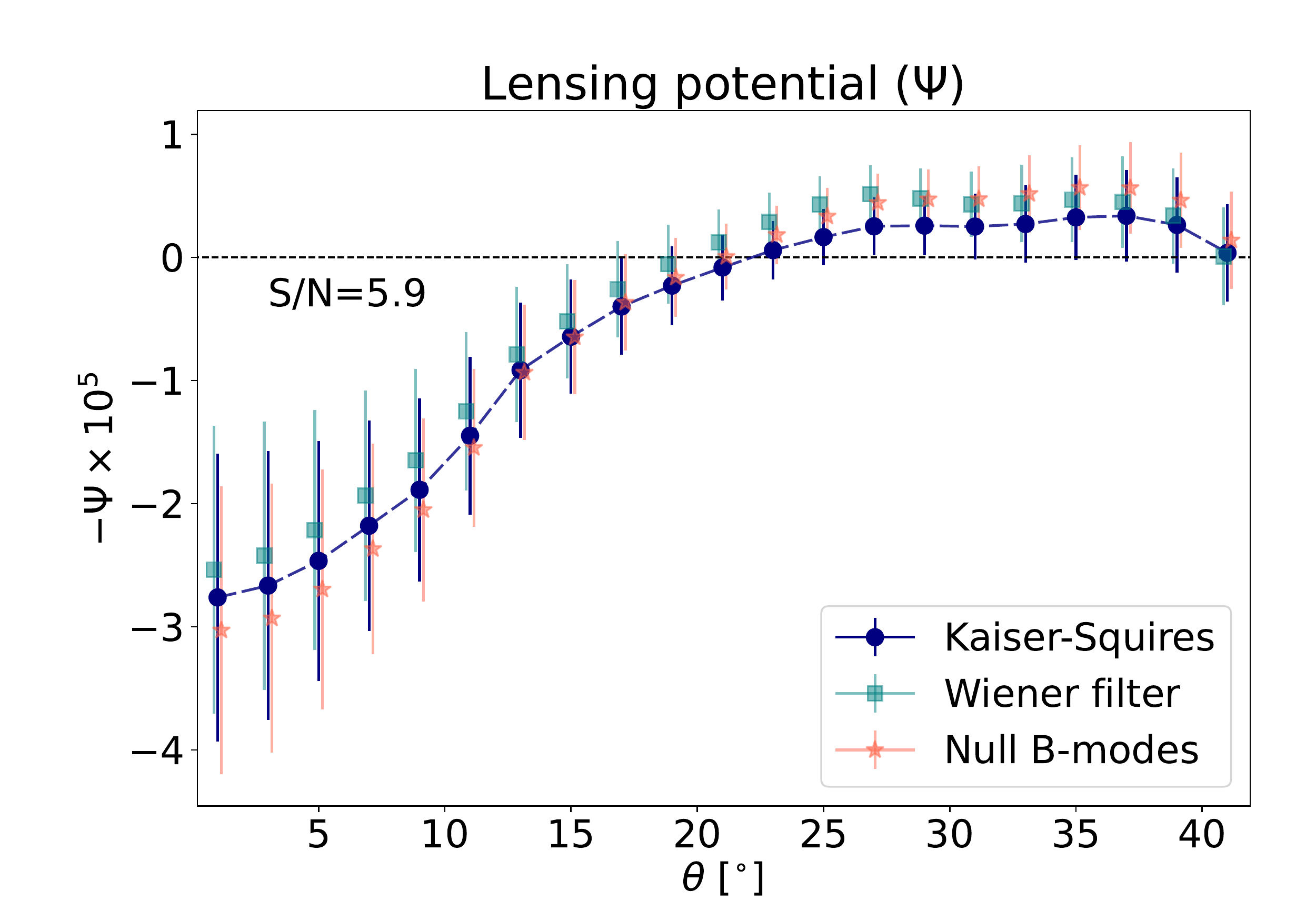}
\end{center}
\caption{Tangential $\psi$ profiles measured from the projected gravitational potential maps that we calculated from the original $\kappa$ maps without smoothing or filtering. We again report consistent results in the full extent of the profile from different mass maps.}
\label{fig:profiles_pot}
\end{figure}

\subsubsection{Lensing potential ($\psi$)}

In Figure \ref{fig:profiles_pot}, we illustrate how the reconstructed large-scale gravitational potential at the CS changes when different map versions are used. As in the previous two cases, we observe a consistent profile shape and amplitude. An under-density is detected with $\mathrm{S/N}\approx2.3$ considering only the first bin, but the overall detection significance reaches $\mathrm{S/N}\approx5.9$ considering the full extent of the profile. Given the errors, the measured profiles from different mass map versions show consistent results, including the compensation zone beyond the zero-crossing at about $\theta\approx20^{\circ}$.

From these tests, we also observed that the transverse size of the under-density extends beyond the actual CS region, and other surrounding voids are also expected to contribute to this large fluctuation in the gravitational potential. We note that another large void with $R\approx250~\mpc$ line-of-sight size, that \cite{Jeffrey2021} detected in alignment with a significant $\kappa<0$ region in the DES Y3 mass map at $\textrm{RA}, \textrm{Dec} \approx 41.2^{\circ}, -12.2^{\circ}$, is also expected to contribute to the gravitational potential fluctuation ($\theta \approx7^{\circ}$ from the CS centre). 

\begin{figure}
\begin{center}
\includegraphics[width=84mm
]{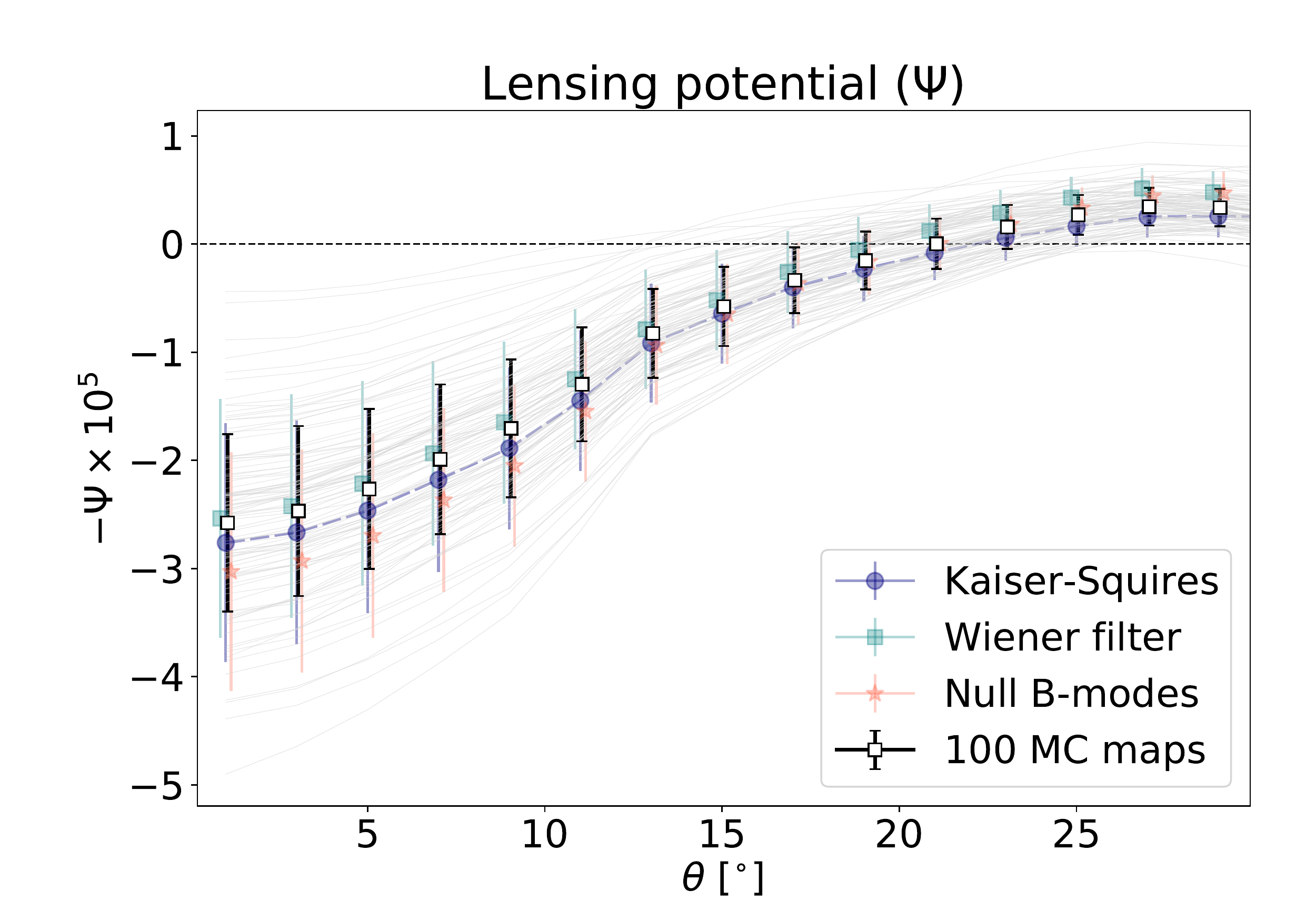}
\end{center}
\caption{Tangential profiles from 100 independent Monte Carlo samples from the posterior $p(\kappa | \gamma)$ using the Wiener posterior (light grey curves) with fixed void position are compared to our main results that are based on maximum a posteriori mass maps from different reconstruction methods. Though the error bars represent different uncertainties, the mean and standard deviation (black markers) of this ensemble is in great agreement with our baseline analyses in the full extent of the profile.}
\label{fig:profiles_pot_MC}
\end{figure}

This finding suggests that the actually deepest part of the Eridanus supervoid may be outside the central CS region, but the overall large-scale fluctuation in the $\psi$ map is very accurately centred on the CS, and this is what drives the expected ISW signal. We leave the more detailed analysis of the substructure of the Eridanus supervoid for future work, possibly including upcoming DES data releases and novel spectroscopic data sets.

\subsection{Monte Carlo sampling}

To evaluate the uncertainty associated with the lensing potential profile at fixed position, we target the posterior probability $p(\psi | \gamma, \rm{RA, Dec})$ given lensing data $\gamma$. We generate independent Monte Carlo (MC) samples from this $\psi$ posterior using samples from the posterior of possible mass maps $p(\kappa | \gamma)$. For each sample of the posterior probability of the map $\kappa$, we evaluate the lensing potential $\psi$ and measure the radial profile for the fixed RA, Dec of the CS. 

For the mass map posterior samples, we use the same Gaussian prior as used by the Wiener filter. The Wiener filter is equivalent to the mean of all possible posterior samples (in addition to being the maximum of the posterior distribution). Each posterior sample is a \textit{constrained realisation}~\citep{cosntrained_realizations,Zaroubi1995} and each appears as a full-sky mass map with the expected power spectrum and masked regions \textit{in painted}. Unobserved regions of the sky are lightly constrained by data and will therefore fluctuate heavily between samples. To generate $\kappa$ realisations drawn from $p(\kappa | \gamma)$ we use the {\sc Dante}\footnote{\url{https://github.com/doogesh/dante}} package~\citep{dante} with settings matching those described by \cite{Jeffrey2021}.

In Figure \ref{fig:profiles_pot_MC}, we present the results from the analysis of 100 MC mass map samples. The uncertainties that we estimated from the MC sampling are comparable to the errors we estimated from randomisation of the void centre in the DES footprint but their actual meaning is slightly different. The MC errors characterise our knowledge on the amplitude and shape of the reconstructed void profile, whereas the errors from randomisation correspond to a detection significance compared to a null hypothesis of no void detection. 

Given these errors, we found that the mean profile calculated from this ensemble of profiles is not identical to the baseline maximum a posteriori Wiener filter map result. However, the two results are fully consistent with each other, and also with the other two mass map versions Kaiser-Squires and Null B-modes.

\section{Void lensing in simulations}
\label{sec:section_nbody}

Considering all the evidence presented, we report a robust $\mathrm{S/N}\gtrsim5$ detection of a supervoid from the DES Y3 mass maps aligned with the CS. Importantly, this finding is fully consistent with the expectation of a $\mathrm{S/N}\gtrsim4$ detection from an Eridanus-like supervoid based on the simulation analysis by \cite{Higuchi2018}. 

\subsection{Methodology}

To better assess the consistency of the observed lensing signal with $\Lambda$CDM expectations, we also analysed convergence maps from N-body simulations. We used a set of full-sky mock lensing maps \citep{Takahashi2017} obtained for source redshifts $z\lesssim1.4$, in consistency with the range of DES Y3 source galaxies. Initial conditions were generated using the \texttt{2LPTIC} code \citep{Crocce2006} and the N-body simulation used \texttt{L-GADGET2} \citep{springel2005} with cosmological parameters consistent with the WMAP 9-year results: $\Omega_{\rm m} = 0.279$, $\sigma_8 = 0.82$, $\Omega_{\rm b} = 0.046$, $n_{\rm s} = 0.97 $, $h = 0.7$ \citep[see][for details]{Hinshaw2013}. 

The average matter power spectra of the simulations agree with the revised \texttt{HALOFIT} \citep{Takahashi2012} predictions within $5$ per cent for $k < 1~h~\mathrm{Mpc}^{-1}$ at $z < 1$ and for $k < 0.8~h~\mathrm{Mpc}^{-1}$ at $z < 3$. A multiple plane ray-tracing algorithm (\texttt{GRayTrix}, \citealt{Hamana2015}) was used to estimate the values of the   convergence fields for the simulation snapshots, and $\kappa$ maps are provided in the form of \texttt{HEALPix} maps with different resolutions, including $N_\mathrm{side} = 4096$ which we used for our tests. 

\begin{figure}
\begin{center}
\includegraphics[width=85mm
]{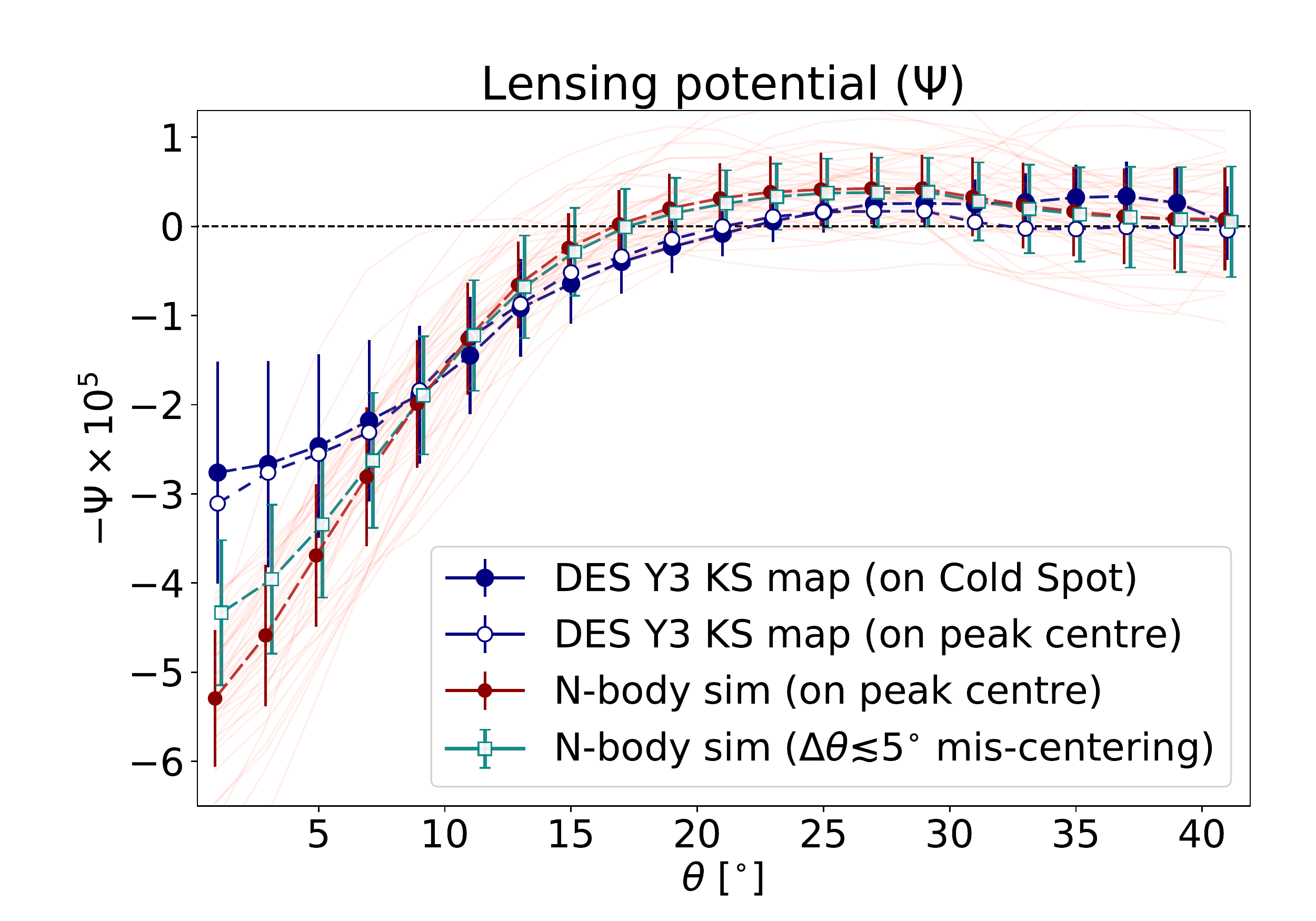}
\end{center}
\caption{A comparison of lensing signals from supervoids, identified as large-scale peaks in $\psi$ maps, using DES Y3 data (dark blue) and 40 mock realisations (pale red curves). The red points and error bars show the mean and standard deviation of the 40 simulated analyses, if the profiles are centred on the actual peak of the potential fluctuations. The discrepancy with observations cannot be resolved even when a random $\Delta\theta\lesssim5^{\circ}$ peak mis-centring is applied in simulations (empty squares). It is also demonstrated that the actual peak in DES Y3 data is very closely aligned with the nominal CS position, and measuring the $\psi$ profile around this observed peak (empty circles) does not significantly change the DES Y3 results.}
\label{fig:profiles_pot_nbody}
\end{figure}

\subsection{Consistency tests}

To model our DES Y3 analysis, we downgraded the mock $\kappa$ maps to $N_\mathrm{side} = 1024$, and converted them to $\psi$ maps by applying the $\kappa_{\ell m}=-\frac{1}{2}\ell(\ell+1)\psi_{\ell m}$ transformation. We considered 5 different mock $\Lambda$CDM realisations, and also analysed each octant separately in the 5 full-sky maps. With this strategy, we accounted for the 4100 deg$^{2}$ survey window of the observed DES data (using a less complicated mask than in observations, but identical map resolutions), and thereby built an ensemble of $5\times8=40$ DES Y3-like simulations to compare to. We also checked that the simulations feature similar fluctuations in the maps compared to the DES data, and that using lower resolution maps does not change the results.

In each of the DES Y3-like octant mocks, we identified the location of the \texttt{HEALPix} pixel with the \emph{highest} value of $\psi$, which corresponds to the approximate centre of the largest projected under-density in the map. As in the DES Y3 analysis, we then measured the tangential profiles around these most significant peaks in the $\psi$ maps, and determined their mean and standard deviation. 

As shown in Figure \ref{fig:profiles_pot_nbody}, we found that, in the centre of the CS ($\theta\lesssim5^{\circ}$), the lensing signal from the Eridanus supervoid is about $30\%$ lower than typical results from the $\Lambda$CDM mocks. The significance of the observed discrepancy is at the $2\sigma$ level, with a $\psi_\mathrm{0}\approx(2.8\pm1.2)\times 10^{-5}$ central value for DES Y3 and $\psi_\mathrm{0}\approx(5.3\pm0.8)\times 10^{-5}$ measured for our mocks. At $\theta\approx3^{\circ}$ and $\theta\approx5^{\circ}$, we also found a $1.7\sigma$ and a $1.2\sigma$ lower-than-expected signal, respectively, followed by consistent signal strength in the rest of the profile. 

We stress that this is a \emph{frequentist} analysis; the cosmological parameters $\Omega_{\rm m} = 0.279$ and $\sigma_8 = 0.82$ are not varied in our mock realisations. Therefore, the $2\sigma$ discrepancy that we found should be interpreted as the likelihood of detecting such a low lensing signal given the specific $\Lambda$CDM model parameters, and also considering the field-to-field fluctuations in the simulated measurements.

\subsection{Discussion $\&$ interpretation}

While this mismatch is not highly significant, we considered possible explanations. First, we tested the expected role of mis-centring in the identification of the largest peak in the $\psi$ maps. As demonstrated in Figure \ref{fig:profiles_pot_nbody}, a randomly assigned $\Delta\theta\lesssim5^{\circ}$ shift of the peak position in mock profile measurements could reduce the $\sim2\sigma$ tension to the even more tolerable $1.3\sigma$ level (and also results in a slight widening of the lensing profiles at large radii, i.e. similar to the DES results). Then, we also identified the position of the large-scale peak in the DES Y3 $\psi$ map, finding $\textrm{RA}, \textrm{Dec} \approx 45.0^{\circ}, -18.5^{\circ}$, i.e. only about $\Delta\theta\approx3.8^{\circ}$ from the nominal CS centre at $\textrm{RA}, \textrm{Dec} \approx 48.3^{\circ}, -20.4^{\circ}$. We measured the tangential profile in the DES Y3 $\psi$ map around this location, and found an approximately $10\%$ stronger signal in the centre (see Figure \ref{fig:profiles_pot_nbody}). Therefore, the observed mis-centring does not explain all of the discrepancy (but it may contribute to it), assuming that the DES mass maps are correct.

We also considered that a lower lensing signal might come from imperfections in the observational reconstruction of the underlying potential $\psi$. The performance of the DES Y3 mass mapping methods was validated using a single realisation of the \cite{Takahashi2017} simulations that we also analysed. In that example, \cite{Jeffrey2021} found that the power spectrum of the reconstructed mass map simulations was under-predicted by $\sim$10-30$\%$ at the largest scales ($\ell\lesssim15$) compared the true noiseless mock convergence field. In a more realistic analysis, a noisy realisation of the same underlying $\kappa$ map was analysed, and no significant bias was found in the KS and Null B-mode map reconstructions. We also highlight here that the DES survey window provides an incomplete map of the surroundings of the wider CS region (see Figure \ref{fig:image_CS}), and thus the reconstruction of the large-scale potential may also be imperfect, both in terms of centring and in overall amplitude.

We note that the large-scale modes may vary quite significantly in different realisations due to cosmic variance. While the role of these variations was not extensively tested in the DES Y3 mass map analyses \citep{Jeffrey2021}, a consistent suppression of large-scale modes in DES Y3 mass map reconstructions could in principle result in lower amplitudes at the large-scale peaks. In our measurements and mock analyses, we made an attempt to minimise such possible biases by setting the $\ell\leq3$ spherical harmonic modes to zero in the $\psi$ map. At the expense of losing some of the signal, this simple high-pass filtering removes super-survey modes from the lensing potential map $\psi$ that a DES Y3-like 4100 deg$^{2}$ survey is unable to probe. In turn, the strongest \emph{local} peaks in the full-sky $\psi$ map's octants became more comparable to the cut-sky DES Y3 maps.

Overall, the flattened lensing profile observed in DES data may well be a \emph{combined} effect from the above sources of imperfections concerning the data (mis-centring, nearby survey edge, suppression of large-scale power in the map reconstructions), but we stress that the observed discrepancy remains at the moderate $2\sigma$ level.

Finally, we also contemplated that the lower-than-expected lensing signal in the direction of the CS centre is due to a genuine physical effect, since there are intriguing precedents for similar findings in cosmology. For example, \cite{Leauthaud2017} reported from their BOSS $\times$ CFHTLenS galaxy-galaxy lensing measurements that the observed lensing signal is $\sim$20-40$\%$ lower than expected based on the auto-correlation of the galaxy sample. Then, \cite{Lange2021} determined that this tension does not significantly depend on the mass of halos in the $10^{13.3}$-$10^{13.9}h^{-1}M_\odot$ range and no significant scale-dependence is seen in the $0.1\mpc<r<60\mpc$ range. These results can exclude some proposed small-scale phenomena as explanations, such as baryonic effects or insufficient halo occupation modelling.

Considering cosmic voids, we highlight that the CMB lensing imprint of DES Y1 voids was also found slightly lower than expected with about $A_{\kappa}=\kappa_\mathrm{obs}/\kappa_\mathrm{th}\approx0.8$ \citep[][]{Vielzeuf2019}. While the significance of this DES Y1 result was only moderate, \cite{Hang2021} analysed a larger sample of similarly defined voids and superclusters using the Legacy Survey photo-$z$ catalogue \citep[][]{Dey2019}, and reported $A_{\kappa}\approx0.811\pm0.057$, i.e. $3.3\sigma$ lower signal than expected from a standard $\Lambda$CDM model. 

As in the case of the ISW excess signals discussed in Section \ref{Section1}, it is important to note that not all void samples show anomalously low lensing signals \citep[see e.g.][]{Cai2017,Raghunathan2019}, and therefore more work is needed to settle this debate, including this analysis of the CS area and the Eridanus supervoid. Taken at face value, a lower-than-expected lensing signal could in principle be a consequence of a faster low-$z$ expansion rate and a related stronger \emph{decay} of the gravitational potentials ($\dot{\Phi}<0$) than assumed in the baseline $\Lambda$CDM model (see Equations \ref{eq:ISW_definition} $\&$ \ref{eq:ISW_definition2}), i.e. low lensing and strong ISW signals are not inconsistent.

We note that the $S_{8}=\sigma_\mathrm{8}\sqrt{\Omega_\mathrm{m}/0.3}\approx0.79$ lensing parameter from the mocks, given the $\Omega_{\rm m} = 0.279$ and $\sigma_8 = 0.82$ parameters, is consistent with the main DES Y3 result $S_{8}\approx0.776\pm0.017$ \citep[][]{descollaboration2021dark}. This implies that while the $\Omega_{\rm m}$ and $\sigma_8$ parameters may differ, the overall lensing amplitude is expected to agree at the $2\%$ level, which provides a good basis for comparison.

\section{Summary $\&$ Conclusions}
\label{sec:section_disc}

In this paper, we investigated the CMB CS region using the Year-3 data set from the Dark Energy Survey \citep[][]{DES_review}. To advance the mapping of the low-$z$ Eridanus supervoid aligned with the CS, we used the \redmagic{} catalogue of LRGs. As a key innovation, we also analysed state-of-the-art weak lensing data in the form of dark matter \emph{mass maps} reconstructed from the DES Y3 data set \citep[][]{Jeffrey2021}.

As demonstrated in Figure \ref{fig:profile_CS}, we first measured the line-of-sight galaxy density profile in the direction of the CS centre. In consistency with previous galaxy surveys \citep[see e.g.][]{Mackenzie2017}, we provided more evidence for the existence of the Eridanus supervoid ($R\approx 200~\mpc$, $\delta_{\rm 0}\approx-0.2$) from the distribution of DES Y3 \redmagic{} galaxies at $z\lesssim0.2$.

We then presented a robust $\mathrm{S/N}\gtrsim5$ detection of the Eridanus supervoid from the reconstructed DES Y3 \emph{mass maps} (see Table \ref{tab:table1}), in line with the expectations from related N-body simulation analyses by \cite{Higuchi2018}. We found no significant difference in the lensing profiles when considering different mass map reconstruction methods (Kaiser-Squires, Wiener filter, Null B-modes). Also, our detection is stable when changing our methodology to highlight the largest scales from the DES $\kappa$ maps ($\sigma=10^{\circ}$ Gaussian smoothing, $\ell<10$ filtering, lensing potential $\psi$). In particular, our findings confirmed that the Eridanus supervoid is the most prominent large-scale under-density in the 4100 deg$^{2}$ survey footprint mapped by the DESY3 data set (see Figure \ref{fig:image_CS}), further suggesting a causal connection with the CS. 

Finally, we tested the lensing signal's amplitude in the direction of the CS. We looked for the strongest large-scale peaks in the lensing potential map $\psi$ in N-body simulations \citep[][]{Takahashi2017}, and compared our DES Y3 results to the resulting ensemble of mock supervoid profiles. Interestingly, we found that the observed lensing imprint of the Eridanus supervoid is $\sim30\%$ \emph{lower} than expected from measurements of the largest voids found in mocks based on the $\Lambda$CDM model. We noted that this discrepancy is observed at the moderate $\sim2\sigma$ significance level (frequentist analysis), restricted only to the CS centre at $\theta\lesssim5^{\circ}$ (see Figure \ref{fig:profiles_pot_nbody} for details).

We nonetheless considered three possible explanations. Using mock lensing potential ($\psi$) maps, we determined that the observed discrepancy cannot be fully resolved by assuming a random $\Delta\theta\lesssim5^{\circ}$ peak mis-centring, but it could reduce the tension to the $1.3\sigma$ level. We also argued that a consistent under-estimation of the large-scale modes in the DES Y3 mass map reconstruction process might also explain such a discrepancy, and this possibility is consistent with existing mass mapping analyses in simulations \citep[][]{Jeffrey2021}. 

As a third option, we also considered that the low lensing signal at the CS is due to a genuine physical effect. We provided examples for $\sim$20-40$\%$ lower-than-expected lensing amplitudes measured from over-densities \citep[][]{Leauthaud2017} and also from voids \citep[][]{Vielzeuf2019,Hang2021}. If the low-$z$ growth rate of structure in supervoids is even more suppressed than in $\Lambda$CDM, that leads to shallower gravitational potentials, weaker lensing effects, and a stronger ISW signal. Note that this interesting possibility is consistent with the excess ISW signals observed from a statistical analysis of $R\gtrsim100~\mpc$ supervoids \citep[][]{Kovacs2019}.

\begin{table}
\centering
\caption{\label{tab:table1} The estimated S/N of a supervoid detection in the KS mass map for the three different map filtering strategies. We compare the detection significances for the most central bin only, and if all 21 bins are used.}
\begin{tabular}{@{}cccc}
\hline
\hline
Map version: & $\psi$ & $\ell < 10$ & $\sigma=10^{\circ}$ \\
\hline
Bins 1-21 & 5.9 & 6.1 & 7.2 \\
Bin 1 only & 2.3 & 2.1 & 2.1
\\
\hline
\hline
\end{tabular}
\end{table}

In the context of the CS, a hitherto unknown alternative cosmological model might also provide explanation for the large enhancement that would be needed to explain its deep $\Delta T_{\rm 0} \approx -150~\mu K$ central temperature depression as an ISW imprint, if the underlying model of dark energy is not the cosmological constant \citep[see e.g.][]{Beck2018, Kovacs2020}. Therefore, possible relations to the Hubble constant tension \citep[see e.g.][]{DiValentino2021} and the $S_8$ problem \citep[see e.g.][]{Heymans2021,Secco2021} should also be explored in greater details. 

We note that not all data sets and methodologies agree on the detection of such excess ISW signals \citep[see e.g.][]{Hang2021}, and the claimed tensions often remain undetected using smaller voids \citep{NadathurCrittenden2016} or two-point correlation functions \citep{Hang20212pt}. Nevertheless, recent ISW measurements using the eBOSS quasar catalog \citep[][]{Ross2020}, covering the $0.8<z<2.2$ range, also showed ISW anomalies at redshifts higher than before \citep[see][for details]{Kovacs2021}, further suggesting an alternative growth history in supervoid environments.

In the light of these findings, the imprint of super-structures in the CMB remains an interesting unsolved problem in cosmology. Future releases of the DES data and other weak lensing and galaxy surveys such as HSC, KiDS, Euclid, eBOSS, and DESI will certainly help to converge to a solution, together with a more precise mapping of the CS including CMB polarisation data \citep[][]{Kang2020}.

\section*{Data availability}

Galaxy catalogs and lensing data will become publicly available as part of the release of the DES Year-3 products\footnote{\url{https://www.darkenergysurvey.org/the-des-project/data-access/}}. The N-body simulations and the corresponding $\kappa$ maps by \cite{Takahashi2017} are publicly available\footnote{\url{http://cosmo.phys.hirosaki-u.ac.jp/takahasi/allsky\textunderscore raytracing/}}. The {\it Planck} CMB temperature maps that we used to visualise the Cold Spot region are also public\footnote{\url{https://www.cosmos.esa.int/web/planck}}.

The data underlying this article including the reconstructed void lensing profiles will be shared on reasonable request to the corresponding author, in line with the relevant DES data policy.

\section*{Acknowledgments}

This paper has gone through \emph{internal review} by the DES collaboration. AK has been supported by a Juan de la Cierva \emph{Incorporaci\'on} fellowship with project number IJC2018-037730-I, and funding for this project was also available in part through SEV-2015-0548 and AYA2017-89891-P. The authors thank an anonymous referee whose insights improved the clarity of the paper.

Funding for the DES Projects has been provided by the U.S. Department of Energy, the U.S. National Science Foundation, the Ministry of Science and Education of Spain, 
the Science and Technology Facilities Council of the United Kingdom, the Higher Education Funding Council for England, the National Center for Supercomputing 
Applications at the University of Illinois at Urbana-Champaign, the Kavli Institute of Cosmological Physics at the University of Chicago, 
the Center for Cosmology and Astro-Particle Physics at the Ohio State University,
the Mitchell Institute for Fundamental Physics and Astronomy at Texas A\&M University, Financiadora de Estudos e Projetos, 
Funda{\c c}{\~a}o Carlos Chagas Filho de Amparo {\`a} Pesquisa do Estado do Rio de Janeiro, Conselho Nacional de Desenvolvimento Cient{\'i}fico e Tecnol{\'o}gico and 
the Minist{\'e}rio da Ci{\^e}ncia, Tecnologia e Inova{\c c}{\~a}o, the Deutsche Forschungsgemeinschaft and the Collaborating Institutions in the Dark Energy Survey. 

The Collaborating Institutions are Argonne National Laboratory, the University of California at Santa Cruz, the University of Cambridge, Centro de Investigaciones Energ{\'e}ticas, 
Medioambientales y Tecnol{\'o}gicas-Madrid, the University of Chicago, University College London, the DES-Brazil Consortium, the University of Edinburgh, 
the Eidgen{\"o}ssische Technische Hochschule (ETH) Z{\"u}rich, 
Fermi National Accelerator Laboratory, the University of Illinois at Urbana-Champaign, the Institut de Ci{\`e}ncies de l'Espai (IEEC/CSIC), 
the Institut de F{\'i}sica d'Altes Energies, Lawrence Berkeley National Laboratory, the Ludwig-Maximilians Universit{\"a}t M{\"u}nchen and the associated Excellence Cluster Universe, 
the University of Michigan, NFS's NOIRLab, the University of Nottingham, The Ohio State University, the University of Pennsylvania, the University of Portsmouth, 
SLAC National Accelerator Laboratory, Stanford University, the University of Sussex, Texas A\&M University, and the OzDES Membership Consortium.

Based in part on observations at Cerro Tololo Inter-American Observatory at NSF's NOIRLab (NOIRLab Prop. ID 2012B-0001; PI: J. Frieman), which is managed by the Association of Universities for Research in Astronomy (AURA) under a cooperative agreement with the National Science Foundation.

The DES data management system is supported by the National Science Foundation under Grant Numbers AST-1138766 and AST-1536171.
The DES participants from Spanish institutions are partially supported by MICINN under grants ESP2017-89838, PGC2018-094773, PGC2018-102021, SEV-2016-0588, SEV-2016-0597, and MDM-2015-0509, some of which include ERDF funds from the European Union. IFAE is partially funded by the CERCA program of the Generalitat de Catalunya.
Research leading to these results has received funding from the European Research
Council under the European Union's Seventh Framework Program (FP7/2007-2013) including ERC grant agreements 240672, 291329, and 306478.
We  acknowledge support from the Brazilian Instituto Nacional de Ci\^encia
e Tecnologia (INCT) do e-Universo (CNPq grant 465376/2014-2).

This manuscript has been authored by Fermi Research Alliance, LLC under Contract No. DE-AC02-07CH11359 with the U.S. Department of Energy, Office of Science, Office of High Energy Physics.

\bibliographystyle{mnras}
\bibliography{refs}

\appendix
\section{Density profile analysis}
\label{appendix1}

\begin{figure*}
\begin{center}
\includegraphics[width=178mm
]{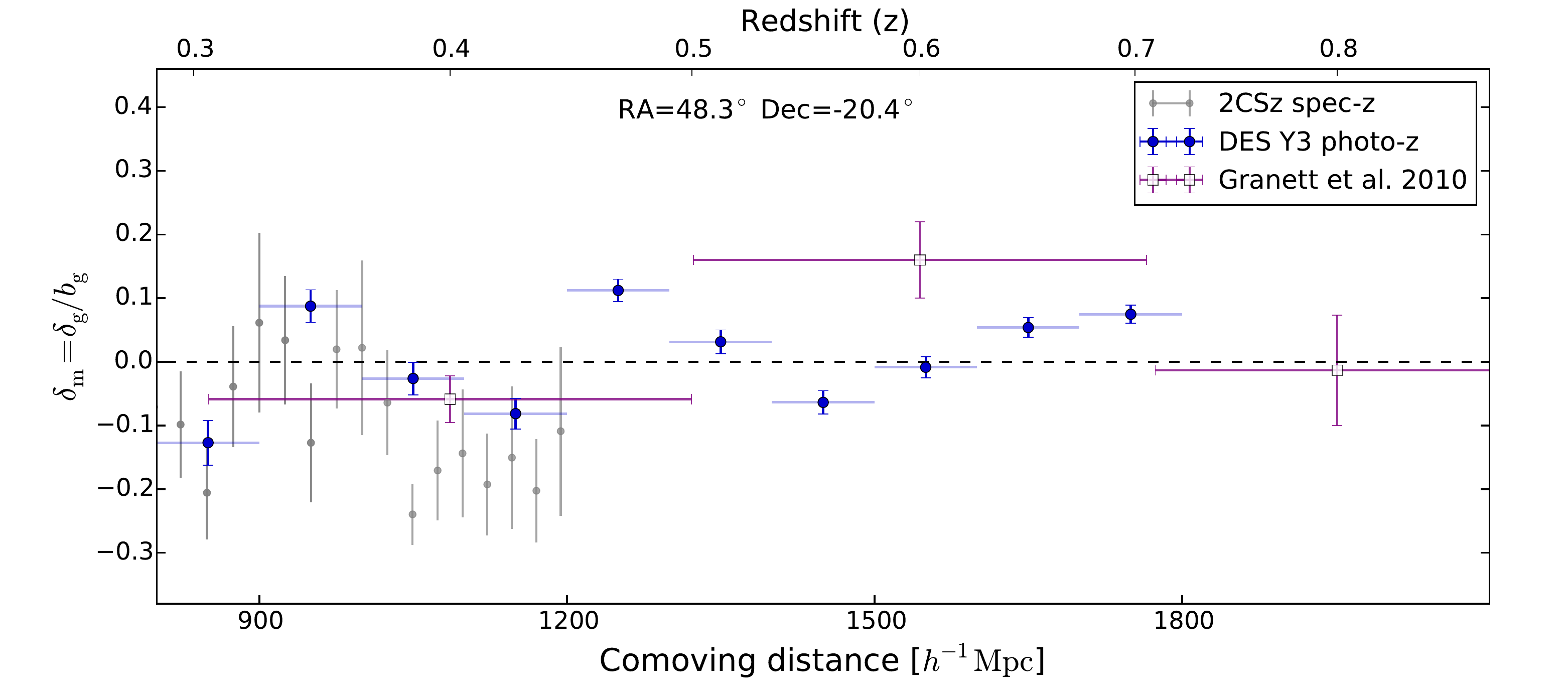}
\end{center}
\caption{Extending the previously shown lower $z$ range, we compare line-of-sight matter density profiles at $0.3<z<0.9$ for different surveys of galaxies in the CS direction. As in Figure \ref{fig:profile_CS}, we converted galaxy density to matter density using the independently determined linear galaxy bias ($b_{\rm g}$) values for each tracer data set. In good consistency with previous results by \citet[]{GranettEtal2010} who conducted a dedicated photo-$z$ survey in the area, the DES results from \redmagic{} galaxies show no evidence for significant voids or over-densities beyond the $z<0.3$ range where the Eridanus supervoid is observed.}
\label{fig:profile_CS2}
\end{figure*}

Here we provide further details on our DES Y3 measurements of the high-$z$ matter density profile in the direction of the CS. In Figure \ref{fig:profile_CS2}, we illustrate that the DES Y3 data shows no evidence for significant voids or over-densities at the $0.3<z<0.9$ range beyond the known under-densities at  $z<0.3$ in Eridanus. These results are consistent the findings by \cite{GranettEtal2010} and \cite{BremerEtal2010}.

\section*{Author Affiliations}
{\small
$^{1}$ Instituto de Astrof\'{\i}sica de Canarias (IAC), Calle V\'{\i}a L\'{a}ctea, E-38200, La Laguna, Tenerife, Spain\\
$^{2}$ Departamento de Astrof\'{\i}sica, Universidad de La Laguna (ULL), E-38206, La Laguna, Tenerife, Spain\\
$^{3}$ Laboratoire de Physique de l'Ecole Normale Sup\'erieure, ENS, Universit\'e PSL, CNRS, Sorbonne Universit\'e, Universit\'e de Paris, Paris, France\\
$^{4}$ Department of Physics \& Astronomy, University College London, Gower Street, London, WC1E 6BT, UK\\
$^{5}$ Institut de F\'{\i}sica d'Altes Energies (IFAE), The Barcelona Institute of Science and Technology, Campus UAB, 08193, Bellaterra, Spain\\
$^{6}$ Department of Physics and Astronomy, University of Pennsylvania, Philadelphia, PA 19104, USA\\
$^{7}$ Department of Astronomy and Astrophysics, University of Chicago, Chicago, IL 60637, USA\\
$^{8}$ Kavli Institute for Cosmological Physics, University of Chicago, Chicago, IL 60637, USA\\
$^{9}$ Universit\"ats-Sternwarte, Fakult\"at f\"ur Physik, Ludwig-Maximilians Universit\"at M\"unchen, Scheinerstr. 1, 81679 M\"unchen, Germany\\
$^{10}$ Institute of Cosmology and Gravitation, University of Portsmouth, Portsmouth, PO1 3FX, UK\\
$^{11}$ Department of Physics, ETH Zurich, Wolfgang-Pauli-Strasse 16, CH-8093 Zurich, Switzerland\\
$^{12}$ Instituto de Fisica Teorica UAM/CSIC, Universidad Autonoma de Madrid, 28049 Madrid, Spain\\
$^{13}$ Argonne National Laboratory, 9700 South Cass Avenue, Lemont, IL 60439, USA\\
$^{14}$ Kavli Institute for Particle Astrophysics \& Cosmology, P. O. Box 2450, Stanford University, Stanford, CA 94305, USA\\
$^{15}$ Physics Department, 2320 Chamberlin Hall, University of Wisconsin-Madison, 1150 University Avenue Madison, WI  53706-1390\\
$^{16}$ Department of Physics, Carnegie Mellon University, Pittsburgh, Pennsylvania 15312, USA\\
$^{17}$ Laborat\'orio Interinstitucional de e-Astronomia - LIneA, Rua Gal. Jos\'e Cristino 77, Rio de Janeiro, RJ - 20921-400, Brazil\\
$^{18}$ Center for Astrophysical Surveys, National Center for Supercomputing Applications, 1205 West Clark St., Urbana, IL 61801, USA\\
$^{19}$ Department of Astronomy, University of Illinois at Urbana-Champaign, 1002 W. Green Street, Urbana, IL 61801, USA\\
$^{20}$ Department of Physics, Duke University Durham, NC 27708, USA\\
$^{21}$ Center for Cosmology and Astro-Particle Physics, The Ohio State University, Columbus, OH 43210, USA\\
$^{22}$ Jodrell Bank Center for Astrophysics, School of Physics and Astronomy, University of Manchester, Oxford Road, Manchester, M13 9PL, UK\\
$^{23}$ Department of Astronomy, University of California, Berkeley,  501 Campbell Hall, Berkeley, CA 94720, USA\\
$^{24}$ Santa Cruz Institute for Particle Physics, Santa Cruz, CA 95064, USA\\
$^{25}$ Fermi National Accelerator Laboratory, P. O. Box 500, Batavia, IL 60510, USA\\
$^{26}$ Department of Physics, The Ohio State University, Columbus, OH 43210, USA\\
$^{27}$ Jet Propulsion Laboratory, California Institute of Technology, 4800 Oak Grove Dr., Pasadena, CA 91109, USA\\
$^{28}$ Department of Physics, Stanford University, 382 Via Pueblo Mall, Stanford, CA 94305, USA\\
$^{29}$ SLAC National Accelerator Laboratory, Menlo Park, CA 94025, USA\\
$^{30}$ Department of Physics, University of Oxford, Denys Wilkinson Building, Keble Road, Oxford OX1 3RH, UK\\
$^{31}$ Department of Astronomy, University of Geneva, ch. d'\'Ecogia 16, CH-1290 Versoix, Switzerland\\
$^{32}$ Department of Physics, University of Michigan, Ann Arbor, MI 48109, USA\\
$^{33}$ Department of Applied Mathematics and Theoretical Physics, University of Cambridge, Cambridge CB3 0WA, UK\\
$^{34}$ Perimeter Institute for Theoretical Physics, 31 Caroline St N, Waterloo, Canada\\
$^{35}$ Instituto de F\'isica Gleb Wataghin, Universidade Estadual de Campinas, 13083-859, Campinas, SP, Brazil\\
$^{36}$ Centro de Investigaciones Energ\'eticas, Medioambientales y Tecnol\'ogicas (CIEMAT), Madrid, Spain\\
$^{37}$ Brookhaven National Laboratory, Bldg 510, Upton, NY 11973, USA\\
$^{38}$ Institut d'Estudis Espacials de Catalunya (IEEC), 08034 Barcelona, Spain\\
$^{39}$ Institute of Space Sciences (ICE, CSIC),  Campus UAB, Carrer de Can Magrans, s/n,  08193 Barcelona, Spain\\
$^{40}$ Max Planck Institute for Extraterrestrial Physics, Giessenbachstrasse, 85748 Garching, Germany\\
$^{41}$ Institute for Astronomy, University of Edinburgh, Edinburgh EH9 3HJ, UK\\
$^{42}$ Departamento de F\'isica Matem\'atica, Instituto de F\'isica, Universidade de S\~ao Paulo, CP 66318, S\~ao Paulo, SP, 05314-970, Brazil\\
$^{43}$ Instituto de F\'{i}sica Te\'orica, Universidade Estadual Paulista, S\~ao Paulo, Brazil\\
$^{44}$ CNRS, UMR 7095, Institut d'Astrophysique de Paris, F-75014, France\\
$^{45}$ Sorbonne Universit\'es, UPMC Univ Paris 06, UMR 7095, Institut d'Astrophysique de Paris, F-75014, Paris, France\\
$^{46}$ Astronomy Unit, Department of Physics, University of Trieste, via Tiepolo 11, I-34131 Trieste, Italy\\
$^{47}$ INAF-Osservatorio Astronomico di Trieste, via G. B. Tiepolo 11, I-34143 Trieste, Italy\\
$^{48}$ Institute for Fundamental Physics of the Universe, Via Beirut 2, 34014 Trieste, Italy\\
$^{49}$ Observat\'orio Nacional, Rua Gal. Jos\'e Cristino 77, Rio de Janeiro, RJ - 20921-400, Brazil\\
$^{50}$ School of Mathematics and Physics, University of Queensland,  Brisbane, QLD 4072, Australia\\
$^{51}$ Department of Physics, IIT Hyderabad, Kandi, Telangana 502285, India\\
$^{52}$ Institute of Theoretical Astrophysics, University of Oslo. P.O. Box 1029 Blindern, NO-0315 Oslo, Norway\\
$^{53}$ Department of Astronomy, University of Michigan, Ann Arbor, MI 48109, USA\\
$^{54}$ Institute of Astronomy, University of Cambridge, Madingley Road, Cambridge CB3 0HA, UK\\
$^{55}$ Kavli Institute for Cosmology, University of Cambridge, Madingley Road, Cambridge CB3 0HA, UK\\
$^{56}$ Center for Astrophysics $\vert$ Harvard \& Smithsonian, 60 Garden Street, Cambridge, MA 02138, USA\\
$^{57}$ Lowell Observatory, 1400 W Mars Hill Rd, Flagstaff, AZ 86001, USA\\
$^{58}$ Australian Astronomical Optics, Faculty of Science and Engineering, Macquarie University, Macquarie Park, NSW 2113, Australia\\
$^{59}$ George P. and Cynthia Woods Mitchell Institute for Fundamental Physics and Astronomy, and Department of Physics and Astronomy, Texas A\&M University, College Station, TX 77843,  USA\\
$^{60}$ Department of Astrophysical Sciences, Princeton University, Peyton Hall, Princeton, NJ 08544, USA\\
$^{61}$ Instituci\'o Catalana de Recerca i Estudis Avan\c{c}ats, Barcelona, Spain\\
$^{62}$ Department of Physics and Astronomy, Pevensey Building, University of Sussex, Brighton, BN1 9QH, UK\\
$^{63}$ School of Physics and Astronomy, University of Southampton,  Southampton, SO17 1BJ, UK\\
$^{64}$ Computer Science and Mathematics Division, Oak Ridge National Laboratory, Oak Ridge, TN 37831\\
}

\end{document}